%% file: main.tex
\documentclass{article}
\usepackage{spconf}
\usepackage{IEEEtrantools}
\usepackage{cite}
\usepackage{hyperref}
\usepackage{graphicx}
\usepackage{tikz}
\usepackage{tikz-3dplot}
    \usetikzlibrary{calc}
    \usetikzlibrary{positioning}
    \usetikzlibrary{arrows}
    \usetikzlibrary{shapes.geometric}
	\usetikzlibrary{arrows.meta}
	\usetikzlibrary{shapes}
\usepackage{pgfplots}
	\usepgfplotslibrary{groupplots}
	\pgfplotsset{compat=newest,every axis/.append style={font=\tiny}}
\usepackage{amsmath,amssymb}
\usepackage{url}
\usepackage{algorithmic}
\usepackage{array}
\usepackage[squaren]{SIunits}
\def\mm{\milli\metre}

\title{Self-Supervised Isotropic Superresolution Fetal Brain MRI}

\name{
    Kay~L\"{a}chler$\dagger\ddagger$, 
    H\'{e}lène~Lajous$\Vert\dagger$,
    Michael~Unser$\ddagger$,
    Meritxell~{Bach Cuadra}$\dagger\Vert$, and 
    Pol~{del Aguila Pla}$\dagger\ddagger$
    }
\address{\small $\dagger$ CIBM Center for Biomedical Imaging, Switzerland \\ 
         \small $\ddagger$ Biomedical Imaging Group, École polytechnique fédérale de Lausanne, Lausanne, Switzerland \\
         \small $\Vert$ Department of Radiology, Lausanne University Hospital (CHUV), and University of Lausanne (UNIL), Lausanne, Switzerland
         }

\input{macros}

\def\titleshift{-6pt}

\begin{document}
    \ninept
    
    \maketitle
    
    \begin{abstract}
        \input{secs/abstract}
    \end{abstract}
    
    \begin{keywords}%
        Image reconstruction, Image enhancement, Neural networks.
    \end{keywords}

    \section{Introduction}
    \label{sec:intro}
        \input{secs/intro}

    \section{Single-Acquisition Isotropic-Resolution Restoration} 
    \label{sec:sair}
        \input{secs/SAIR}

    \section{Data Sets and Evaluation}
    \label{sec:data}
        \input{secs/data}

    \section{Empirical Results} 
    \label{sec:empirical_results}
        \input{secs/empirical_results}

    \section{Discussion}
    \label{sec:Discussion}
        \input{secs/discussion}

    \section{Compliance with Ethical Standards}
    \label{sec:ethics}
    
        The use of the clinical MRI dataset was reviewed and approved by Commission cantonale (VD) d'éthique de la recherche sur l'être humain (CER-VD). Written informed consent for participation was not required for this study, in accordance with the national legislation and the institutional requirements.
    
    \section{Acknowledgements}
    \label{sec:ack}
    
        This work was supported by the Swiss National Science Foundation
        (205321-182602). We acknowledge access to the facilities and
        expertise of the CIBM Center for Biomedical Imaging, a Swiss 
        research center of excellence founded and supported by Lausanne
        University Hospital (CHUV), University of Lausanne (UNIL), École
        polytechnique fédérale de Lausanne (EPFL), University of Geneva
        (UNIGE), and Geneva University Hospitals (HUG).

\bibliographystyle{IEEEbib}
\bibliography{spline_stuff.bib}

\end{document}

%% file: macros.tex
\def\lowresvol{\mathbf{X}^{\mathrm{LR}}} 
\def\highresvol{\mathbf{X}}
\def\upsampledvol{\mathbf{X}^{\mathrm{up}}}
\def\predvol{\mathbf{X}^{\mathrm{pred}}}
\def\blur{\mathbf{B}}
\def\rotblur{\tilde{\blur}}
\newcommand{\downsampling}[1]{\mathbf{D}_{#1}}
\newcommand{\upsampling}[1]{\mathbf{U}_{#1}}
\def\noise{\mathbf{n}}

\newcommand{\rotationang}[1]{\mathbf{R}_{#1}}
\def\rotation{\rotationang{\theta}}

%% file: secs/abstract.tex
Superresolution T2-weighted fetal-brain magnetic-resonance imaging (FBMRI) traditionally
relies on the availability of several orthogonal low-resolution series of 2-dimensional
thick slices (volumes). In practice, only a few low-resolution volumes are acquired. Thus,
optimization-based image-reconstruction methods require strong regularization using hand-crafted
regularizers (\emph{e.g.}, TV). 
Yet, due to \textit{in utero} fetal motion and the rapidly 
changing fetal brain anatomy, the acquisition of the high-resolution images that are required to train supervised learning
methods is difficult. In this paper, we sidestep this difficulty by providing a proof of concept of a self-supervised single-volume 
superresolution framework for T2-weighted FBMRI (SAIR). We validate SAIR quantitatively
in a motion-free simulated environment. Our results for different noise levels and resolution 
ratios suggest that SAIR is comparable to multiple-volume superresolution reconstruction methods. 
We also evaluate SAIR qualitatively on clinical FBMRI data. The results suggest 
SAIR could be incorporated into current reconstruction pipelines.

%% file: secs/intro.tex
Magnetic resonance imaging (MRI) of the \textit{in utero} developing brain is a key
complementary imaging tool to ultrasound, due to its proven ability to visualize
the fetal anatomy with excellent soft tissue contrast. T2-weighted (T2w) imaging is
the tool of choice to depict normal and pathological 
maturation in fetal MRI~\cite{garel_mri_2004,prayer_mri_2006}. Unpredictable fetal motion in the 
womb is a major challenge and leads to the imperative requirement of fast clinical 
acquisitions. In practice, ultra fast multi-slice single-shot sequences are acquired,
resulting in several orthogonal low-resolution (LR) series (also referred to as volumes) during
one session. At \unit{1.5}{\tesla}, these volumes have a very good in-plane spatial 
resolution ($\Delta_\mathrm{x}=\Delta_\mathrm{y}\approx\unit{1}{\mm}$) while the slice thickness 
is typically chosen between $\Delta_\mathrm{z}=\unit{3}{\mm}$ and $\unit{5}{\mm}$ 
in order to ensure a good signal-to-noise ratio. This results in a resolution ratio 
$r=\Delta_\mathrm{z}/\Delta_\mathrm{x}\in[3,5]$. 

\begin{figure}
    \centering
    \resizebox{0.40\textwidth}{!}{\input{figs/SAIR_diagram_training.tikz}}

    \vspace{-5pt}
    
    \caption{Training procedure for 
        the (SAIR) reconstruction pipeline. \label{fig:SAIR_train}}
\end{figure}
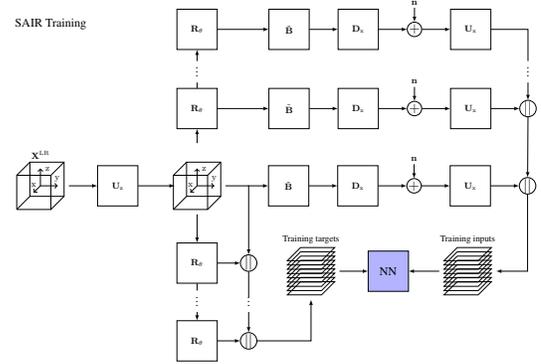

Several methods have been proposed for the superresolution (SR) reconstruction of fetal T2w 
imaging~\cite{Rousseau2006,Gholipour2010,Murgasova2012,Fogtmann2012,art:Tourbier2015,kainz2015fast,cardoso_context-sensitive_2017,ebner_automated_2020,song_deep_2022,xu_svort_2022}.
These support fetal-brain explorations and allow for the automated quantitative study of isotropic high resolution (HR)
3-dimensional (3D) volumes~\cite{makropoulos_review_2018}. Most fetal SR algorithms use multiple orthogonal LR volumes to reconstruct a single isotropic
HR volume. 
Traditional approaches are based on
iterative optimization schemes, which combine motion-estimation and 
image-reconstruction~\cite{Rousseau2006,Gholipour2010,Murgasova2012,Fogtmann2012,art:Tourbier2015,kainz2015fast} steps. 
In the motion-estimation step, each volume is aligned to a single reference volume using volume-to-volume registration,
and each slice within each volume is aligned using slice-to-volume registration. The image enhancement is then 
formulated as an ill-posed inverse problem with hand crafted regularizers (\emph{e.g.}, Tikhonov, total-variation, or non local means).
Prior to the SR reconstruction, these methods need to perform a fetal-brain segmentation and a bias-field
correction~\cite{uus_retrospective_2022}. 

Deep-learning SR methods are widely explored in MRI. Despite this, their application to fetal-brain MRI (FBMRI) is 
very limited. Few deep-learning strategies have been proposed within the context of multiple-volume SR reconstruction
of the fetal brain~\cite{cardoso_context-sensitive_2017,song_deep_2022,xu_svort_2022}. 
Even then, they tackle the SR problem only partially, either by providing an initial step to optimization-based approaches~\cite{cardoso_context-sensitive_2017,xu_svort_2022}, or by improving only the in-plane 
(but not the trough-plane) spatial resolution~\cite{song_deep_2022}. In this paper, we propose a method that uses 
a single LR volume and that was designed to obtain a volume with isotropic resolution. 

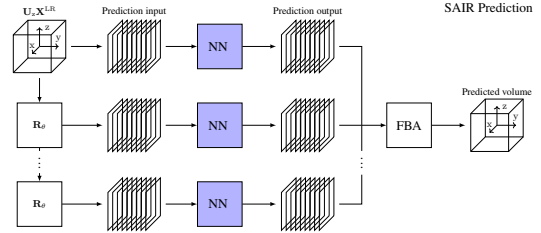
\begin{figure}
    \centering
    \resizebox{0.40\textwidth}{!}{\input{figs/SAIR_diagram_prediction.tikz}}
    
    \vspace{-5pt}
    
    \caption{Schematic description of the prediction procedure for 
        the SAIR reconstruction pipeline. \label{fig:SAIR_pred}}
\end{figure}

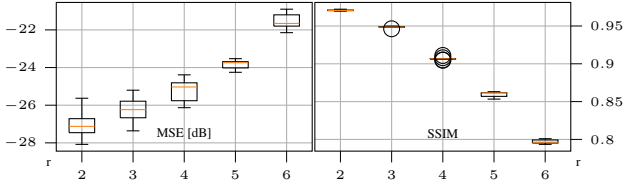
\begin{figure}
    \centering
    \input{figs/R_study}
    
    \vspace{-10pt}
    
    \caption{Performance of the SAIR pipeline when applied to data generated by the simplified MRI
        model~\eqref{eq:simplified_model} applied to the HR volume described in Section~\ref{sec:fabian_data}. Results are 
        reported for nine noise realizations in terms of the resolution ratio $r$. \label{fig:different_rs}}
\end{figure}

\subsection{Related Work on Single-Volume Superresolution}

To the best of our knowledge, only two  methods have been presented for single-volume SR in diffusion and dynamic functional fetal MRI~\cite{kebiri2022through,xu_stress_2021}---but none for T2w imaging. In~\cite{kebiri2022through}, an auto-encoder model is proposed to leverage HR diffusion-weighted 
MRI images from preterm babies to enhance the spatial resolution of a fetal diffusion MRI acquisition. 
This approach relies on training data from preexisting HR datasets, which are hard to obtain and come 
from a different population. To avoid the need for HR data, self-supervised SR approaches like the one we propose  use internal information from the LR image to train the reconstruction procedure. 

In fluorescence microscopy, Weigert \emph{et al.}~\cite{weigert_isotropic_2017} proposed a SR reconstruction method where the information
along the LR axis (by convention, $\mathrm{z}$) is learned from the other two HR axes ($\mathrm{x}$ and $\mathrm{y}$). The distortion
in the LR axis was modeled mathematically and synthetically applied along one of the HR axes to construct pairs of
samples to train a network. The resulting network was used to enhance the resolution along the LR axis. 
The work by Weigert \emph{et al.} inspired our work, which transitions its key ideas to MRI.
Our pipeline is similar to the one independently developed in~\cite{zhao_smore_2021}. Our network, however,
has much fewer parameters and builds on a simpler MRI model. Furthermore, our quantitative empirical study 
avoids the inverse crime by using a realistic physics-based MR-acquisition simulator~\cite{Lajous2022}. We also 
provide a thorough comparison with multiple-volume SR reconstruction methods, which is critical for FBMRI.
In~\cite{xu_stress_2021}, a method that extends the same concepts to both the spatial and temporal domains is 
presented for dynamic functional fetal MRI. 

\subsection{Contribution}

In this paper, we build on~\cite{weigert_isotropic_2017} to develop a single-volume self-supervised SR method for
structural T2w FBMRI.
Such methods can be of very high practical value in FBMRI. First,
they avoid the need for the harmonization of intensities between the LR series, which is a major pitfall for 
quantitative applications with current multiple-volume approaches. Second, they can have a significant impact
on T2 mapping strategies~\cite{lajous_t2_2020} by significantly reducing the acquisition time. Currently,
these strategies can increase the scanning time by up to \unit{12}{\minute}, in the case of the acquisition of
three (or more) orthogonal series at six different echo times.

\begin{figure}
    \centering
    \input{figs/fabian_nomotion_1_inverse_crime.tex}
    
    \vspace{-5pt}
    
    \caption{Left panel: Reconstruction for $r=4$ as described in Section~\ref{sec:results_simplified-model}. The results of b) SAIR reconstruction using a) one single LR series is compared to a synthetic 3D isotropic high resolution ground-truth image. The SSIM is reported in the bottom row.
    Right panel: Zoom over regions of interest. 
    \label{fig:r4-example}}
\end{figure}
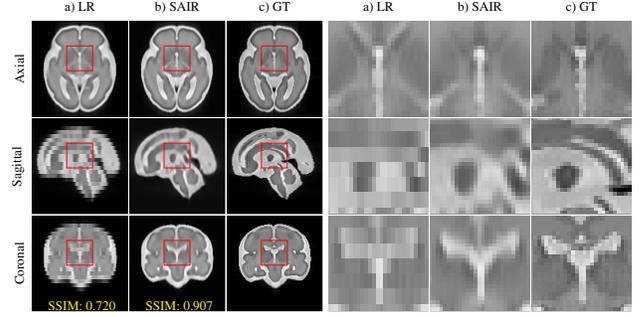

%% file: figs/SAIR_diagram_training.tikz
\newcommand{\VOLUME}[4]{
    \coordinate (fsw) at (#1-#4,#2-#4,#3-#4);
    \coordinate (fse) at (#1+#4,#2-#4,#3-#4);
    \coordinate (fnw) at (#1-#4,#2+#4,#3-#4);
    \coordinate (fne) at (#1+#4,#2+#4,#3-#4);
    \coordinate (bsw) at (#1-#4,#2-#4,#3+#4);
    \coordinate (bse) at (#1+#4,#2-#4,#3+#4);
    \coordinate (bnw) at (#1-#4,#2+#4,#3+#4);
    \coordinate (bne) at (#1+#4,#2+#4,#3+#4);
    \draw (fsw) -- (fse) -- (fne) -- (fnw) -- (fsw);
    \draw (bsw) -- (bse) -- (bne) -- (bnw) -- (bsw);
    \draw (fsw) -- (bsw);
    \draw (fse) -- (bse);
    \draw (fnw) -- (bnw);
    \draw (fne) -- (bne);
    \coordinate (coord_center) at (#1,#2,#3);
    \draw[->] (coord_center) -- +(#4,0,0);
    \draw[->] (coord_center) -- +(0,#4,0);
    \draw[->] (coord_center) -- +(0,0,#4);
    \node (x) at (#1,#2+0.25,#3+#4) {$\mathrm{x}$};
    \node (y) at (#1+#4,#2+0.25,#3) {$\mathrm{y}$};
    \node (z) at (#1+0.25,#2+#4,#3) {$\mathrm{z}$};
}

\newcommand{\xySlices}[2]{
    \foreach \i in {-0.6, -0.45, -0.3, -0.15, 0, 0.15, 0.3, 0.45, 0.6}{
        \draw (-0.6+#1,\i-#2,-0.6) -- (0.6+#1,\i-#2,-0.6) -- (0.6+#1,\i-#2,0.6) -- (-0.6+#1,\i-#2,0.6) -- (-0.6+#1,\i-#2,-0.6);
    }
}

\tikzstyle{square}=[draw, regular polygon, regular polygon sides=4, inner sep=0, minimum size=2cm, align=center]

\tikzstyle{arrow}=[-latex, thick]

\begin{tikzpicture}

\node [anchor=north west] at (-1,6) {\large SAIR Training};

\coordinate (start) at (0,0,0);
\VOLUME{0}{0}{0}{0.6}
\node[above=0.85cm of start] (GT) {$\lowresvol$};
\node[square, right=2cm of start] (Uz) {$\upsampling{\mathrm{z}}$};
\coordinate (upsampled) at (5.5,0,0);
\VOLUME{5.5}{0}{0}{0.6}

\node[below=2cm of upsampled, square] (rot_1) {$\rotation$};
\node[right=0.8cm of rot_1, align=center, circle, draw, inner sep=0] (cat_1) {\Large{$||$}};
\node[below=0.25cm of rot_1, align=center, inner sep=0] (cont) {\large{$\vdots$}};
\node[align=center, inner sep=0] (cont_cat) at (cont -| cat_1) {\large{$\vdots$}};
\node[below=0.5cm of cont, square] (rot_2) {$\rotation$};
\node[align=center, circle, draw, inner sep=0] (cat_2) at (rot_2 -| cat_1) {\Large{$||$}};
\coordinate (targets) at (9.5,-3,0);
\xySlices{9.5}{3}
\node[above=0.85cm of targets] (targets_text) {Training targets};

\coordinate (input) at (15,-3,0);
\xySlices{15}{3}
\node[above=0.85cm of input] (inputs_text) {Training inputs};
\node[right=2cm of targets, align=center, square, fill=blue!30] (NN) {\large{NN}};

\node[right=2.5cm of upsampled, square] (h_0) {$\Tilde{\blur}$};
\node[right=1cm of h_0, square] (Dx_0) {$\downsampling{\mathrm{x}}$};
\node[right=1cm of Dx_0, align=center, circle, draw, inner sep=0] (add_0) {\Large{$+$}};
\node[above=.5cm of add_0, align=center] (noise_0) {$\noise$};
\node[right=1cm of add_0, square] (Ux_0) {$\upsampling{\mathrm{x}}$};
\node[right=1cm of Ux_0, align=center, circle, draw, inner sep=0] (cat_0_i) {\Large{$||$}};
\node[above=2cm of upsampled, square] (rot_1_i) {$\rotation$};
\node[right=1.8cm of rot_1_i, square] (h_1) {$\Tilde{\blur}$};
\node[right=1cm of h_1, square] (Dx_1) {$\downsampling{\mathrm{x}}$};
\node[right=1cm of Dx_1, align=center, circle, draw, inner sep=0] (add_1) {\Large{$+$}};
\node[above=.5cm of add_1, align=center] (noise_1) {$\noise$};
\node[right=1cm of add_1, square] (Ux_1) {$\upsampling{\mathrm{x}}$};
\node[right=1cm of Ux_1, align=center, circle, draw, inner sep=0] (cat_1_i) {\Large{$||$}};
\node[above=0.4cm of rot_1_i, align=center, inner sep=0] (cont_i) {\large{$\vdots$}};
\node[align=center, inner sep=0] (cont_cat_i) at (cont_i -| cat_1_i) {\large{$\vdots$}};
\node[above=0.4cm of cont_i, square] (rot_2_i) {$\rotation$};
\node[right=1.8cm of rot_2_i, square] (h_2) {$\Tilde{\blur}$};
\node[right=1cm of h_2, square] (Dx_2) {$\downsampling{\mathrm{x}}$};
\node[right=1cm of Dx_2, align=center, circle, draw, inner sep=0] (add_2) {\Large{$+$}};
\node[above=.5cm of add_2, align=center] (noise_2) {$\noise$};
\node[right=1cm of add_2, square] (Ux_2) {$\upsampling{\mathrm{x}}$};

\draw[arrow] (1,0) -- (Uz);
\draw[arrow] (Uz) -- +(2, 0);
\draw[arrow] (upsampled)+(0, -1) -- (rot_1);
\draw[arrow] (upsampled)+(1, 0) -| (cat_1);
\draw[thick] (rot_1.south) -- (cont);
\draw[arrow] (rot_1.east) -- (cat_1);
\draw[arrow] (cont.south)+(0,-0.2) -- (rot_2);
\draw[arrow] (rot_2.east) -- (cat_2);
\draw[thick] (cat_1.south) -- +(0,-0.9);
\draw[arrow] (cont_cat.south)+(0,-0.2) -- (cat_2);
\draw[arrow] (cat_2.east) -| ($(targets)+(0,-1)$);

\draw[arrow] (upsampled)+(1.8, 0) -- (h_0);
\draw[arrow] (h_0.east) -- (Dx_0);
\draw[arrow] (Dx_0.east) -- (add_0);
\draw[arrow] (add_0.east) -- (Ux_0);
\draw[arrow] (noise_0.south) -- (add_0);
\draw[arrow] (Ux_0.east) -- (cat_0_i);
\draw[arrow] (upsampled)+(0, +1.5) -- (rot_1_i);
\draw[arrow] (rot_1_i.east) -- (h_1);
\draw[arrow] (h_1.east) -- (Dx_1);
\draw[arrow] (Dx_1.east) -- (add_1);
\draw[arrow] (add_1.east) -- (Ux_1);
\draw[arrow] (noise_1.south) -- (add_1);
\draw[arrow] (Ux_1.east) -- (cat_1_i);

\draw[thick] (rot_1_i) -- ($(cont_i.south)+(0,-0.2)$);
\draw[arrow] (cont_i.north) -- (rot_2_i);
\draw[arrow] (rot_2_i.east) -- (h_2);
\draw[arrow] (h_2.east) -- (Dx_2);
\draw[arrow] (Dx_2.east) -- (add_2);
\draw[arrow] (add_2.east) -- (Ux_2);
\draw[arrow] (noise_2.south) -- (add_2);
\draw[thick] (Ux_2.east) -| (cont_cat_i);
\draw[arrow] (cont_cat_i.south)+(0,-0.2) -- (cat_1_i);
\draw[arrow] (cat_1_i.south) -- (cat_0_i);
\draw[arrow] (cat_0_i.south) |- ($(input)+(1,0)$);

\draw[arrow] (targets)+(1,0) -- (NN);
\draw[arrow] (input)+(-1,0) -- (NN);

\end{tikzpicture}

%% file: figs/SAIR_diagram_prediction.tikz
\newcommand{\VOLUME}[4]{
    \coordinate (fsw) at (#1-#4,#2-#4,#3-#4);
    \coordinate (fse) at (#1+#4,#2-#4,#3-#4);
    \coordinate (fnw) at (#1-#4,#2+#4,#3-#4);
    \coordinate (fne) at (#1+#4,#2+#4,#3-#4);
    \coordinate (bsw) at (#1-#4,#2-#4,#3+#4);
    \coordinate (bse) at (#1+#4,#2-#4,#3+#4);
    \coordinate (bnw) at (#1-#4,#2+#4,#3+#4);
    \coordinate (bne) at (#1+#4,#2+#4,#3+#4);
    \draw (fsw) -- (fse) -- (fne) -- (fnw) -- (fsw);
    \draw (bsw) -- (bse) -- (bne) -- (bnw) -- (bsw);
    \draw (fsw) -- (bsw);
    \draw (fse) -- (bse);
    \draw (fnw) -- (bnw);
    \draw (fne) -- (bne);
    \coordinate (coord_center) at (#1,#2,#3);
    \draw[->] (coord_center) -- +(#4,0,0);
    \draw[->] (coord_center) -- +(0,#4,0);
    \draw[->] (coord_center) -- +(0,0,#4);
    \node (x) at (#1,#2+0.25,#3+#4) {$\mathrm{x}$};
    \node (y) at (#1+#4,#2+0.25,#3) {$\mathrm{y}$};
    \node (z) at (#1+0.25,#2+#4,#3) {$\mathrm{z}$};
}

\newcommand{\xzSlices}[2]{
    \foreach \i in {-0.6, -0.45, -0.3, -0.15, 0, 0.15, 0.3, 0.45, 0.6}{
        \draw (\i+#1,-0.6+#2,-0.6) -- (\i+#1,0.6+#2,-0.6) -- (\i+#1,0.6+#2,0.6) -- (\i+#1,-0.6+#2,0.6) -- (\i+#1,-0.6+#2,-0.6);
    }
}

\tikzstyle{square}=[draw, regular polygon, regular polygon sides=4, inner sep=0, minimum size=2cm, align=center]

\tikzstyle{arrow}=[-latex, thick]

\begin{tikzpicture}

\node [anchor=north east] at (15.75,1.5) {\large SAIR Prediction};

\coordinate (start) at (0,0,0);
\coordinate (upsampled) at (0,0,0);
\VOLUME{0}{0}{0}{0.6}
\node[above=0.85cm of upsampled] (upsampled_text) {{$\upsampling{\mathrm{z}} \lowresvol$}};
\node[square] (rot_1) at ($(upsampled)+(0,-2.5)$) {{$\rotation$}};
\node[below=0.1cm of rot_1, align=center, inner sep=0] (cont) {\large{$\vdots$}};
\node[square] (rot_2) at ($(rot_1)+(0,-2.5)$) {{$\rotation$}};
\coordinate (input) at (3,0,0);
\xzSlices{3}{0}
\node[above=0.85cm of input] (targets_text) {{Prediction input}};
\coordinate (input_1) at (3,-2.5,0);
\xzSlices{3}{-2.5}
\coordinate (input_2) at (3,-5,0);
\xzSlices{3}{-5}
\coordinate (output) at (8.5,0,0);
\xzSlices{8.5}{0}
\node[above=0.85cm of output] (output_text) {{Prediction output}};
\coordinate (output_1) at (8.5,-2.5,0);
\xzSlices{8.5}{-2.5}
\coordinate (output_2) at (8.5,-5,0);
\xzSlices{8.5}{-5}
\node[right=2cm of input, align=center, square, fill=blue!30] (NN) {\large{NN} };
\node[right=2cm of input_1, align=center, square, fill=blue!30] (NN_1) {\large{NN} };
\node[right=2cm of input_2, align=center, square, fill=blue!30] (NN_2) {\large{NN} };
\node[right=2.5cm of output_1, square] (FBA) {\large{FBA}};
\coordinate (FBA_node) at ($(FBA)+(-1.5,0)$);
\node[align=center, inner sep=0] (cont_2) at (cont -| FBA_node) {\large{$\vdots$}};
\coordinate (predicted) at (14.5,-2.5,0);
\VOLUME{14.5}{-2.5}{0}{0.6}
\node[above=0.85cm of predicted, align=center] (predicted_text) {{Predicted volume}};

\draw[arrow] (upsampled)+(0, -1) -- (rot_1);
\draw[arrow] (upsampled)+(1,0) -- +(2, 0);
\draw[thick] (rot_1.south) -- (cont);
\draw[arrow] (cont.south)+(0,-0.15) -- (rot_2);
\draw[arrow] (input)+(1,0) -- (NN);
\draw[arrow] (NN) -- ($(output)+(-1,0)$);
\draw[arrow] (rot_1.east) -- ($(input_1)+(-1,0)$);
\draw[arrow] (input_1)+(1,0) -- (NN_1);
\draw[arrow] (NN_1) -- ($(output_1)+(-1,0)$);
\draw[arrow] (rot_2.east) -- ($(input_2)+(-1,0)$);
\draw[arrow] (input_2)+(1,0) -- (NN_2);
\draw[arrow] (NN_2) -- ($(output_2)+(-1,0)$);
\draw[thick] (output)+(1,0) -| (FBA_node);
\draw[thick] (output_1)+(1,0) -- (FBA_node);
\draw[thick] (output_2)+(1,0) -| ($(cont_2.south)+(0,-0.2)$);
\draw[thick] (cont_2.north) -- (FBA_node);
\draw[arrow] (FBA_node) -- (FBA);
\draw[arrow] (FBA.east) -- ($(predicted)+(-1,0)$);

\end{tikzpicture}

%% file: figs/R_study.tex
\def\hs{1pt}
\def\vs{1pt}
\begin{tikzpicture}

\definecolor{darkgray176}{RGB}{176,176,176}
\definecolor{darkorange25512714}{RGB}{255,127,14}

\begin{groupplot}[group style={group size=2 by 1,horizontal sep=\hs, vertical sep=\vs},height={.2\textwidth}, width={.28\textwidth},grid=both,title style={anchor=center,at={(0.5,0.01)}}]
\nextgroupplot[
xlabel={r},
x label style={at={(-0.03,0)}},
tick align=outside,
tick pos=left,
title={MSE [dB]},
x grid style={darkgray176},
xmin=1.5, xmax=6.5,
xtick={2,3,...,6},
xtick style={color=black},
y grid style={darkgray176},
ylabel={},
ymin=-28.4392491252441, ymax=-20.5414081193347,
ytick style={color=black},
]
\addplot [black]
table {%
1.75 -27.4587577532672
2.25 -27.4587577532672
2.25 -26.714645178681
1.75 -26.714645178681
1.75 -27.4587577532672
};
\addplot [black]
table {%
2 -27.4587577532672
2 -28.0802563522482
};
\addplot [black]
table {%
2 -26.714645178681
2 -25.6305842192748
};
\addplot [black]
table {%
1.875 -28.0802563522482
2.125 -28.0802563522482
};
\addplot [black]
table {%
1.875 -25.6305842192748
2.125 -25.6305842192748
};
\addplot [black]
table {%
2.75 -26.6669755831393
3.25 -26.6669755831393
3.25 -25.7874915846363
2.75 -25.7874915846363
2.75 -26.6669755831393
};
\addplot [black]
table {%
3 -26.6669755831393
3 -27.3609346997136
};
\addplot [black]
table {%
3 -25.7874915846363
3 -25.2028639369079
};
\addplot [black]
table {%
2.875 -27.3609346997136
3.125 -27.3609346997136
};
\addplot [black]
table {%
2.875 -25.2028639369079
3.125 -25.2028639369079
};
\addplot [black]
table {%
3.75 -25.7607785862144
4.25 -25.7607785862144
4.25 -24.8106397497414
3.75 -24.8106397497414
3.75 -25.7607785862144
};
\addplot [black]
table {%
4 -25.7607785862144
4 -26.1316258556936
};
\addplot [black]
table {%
4 -24.8106397497414
4 -24.3873388622281
};
\addplot [black]
table {%
3.875 -26.1316258556936
4.125 -26.1316258556936
};
\addplot [black]
table {%
3.875 -24.3873388622281
4.125 -24.3873388622281
};
\addplot [black]
table {%
4.75 -24.0188849070512
5.25 -24.0188849070512
5.25 -23.6877363205554
4.75 -23.6877363205554
4.75 -24.0188849070512
};
\addplot [black]
table {%
5 -24.0188849070512
5 -24.2521630857789
};
\addplot [black]
table {%
5 -23.6877363205554
5 -23.5288629282272
};
\addplot [black]
table {%
4.875 -24.2521630857789
5.125 -24.2521630857789
};
\addplot [black]
table {%
4.875 -23.5288629282272
5.125 -23.5288629282272
};
\addplot [black]
table {%
5.75 -21.7993389051016
6.25 -21.7993389051016
6.25 -21.1997842246393
5.75 -21.1997842246393
5.75 -21.7993389051016
};
\addplot [black]
table {%
6 -21.7993389051016
6 -22.1469860812554
};
\addplot [black]
table {%
6 -21.1997842246393
6 -20.9004008923306
};
\addplot [black]
table {%
5.875 -22.1469860812554
6.125 -22.1469860812554
};
\addplot [black]
table {%
5.875 -20.9004008923306
6.125 -20.9004008923306
};
\addplot [darkorange25512714]
table {%
1.75 -27.1239973896748
2.25 -27.1239973896748
};
\addplot [darkorange25512714]
table {%
2.75 -26.2347864205005
3.25 -26.2347864205005
};
\addplot [darkorange25512714]
table {%
3.75 -25.0312893123852
4.25 -25.0312893123852
};
\addplot [darkorange25512714]
table {%
4.75 -23.741141997861
5.25 -23.741141997861
};
\addplot [darkorange25512714]
table {%
5.75 -21.6567065701246
6.25 -21.6567065701246
};

\nextgroupplot[
xlabel={r},
x label style={at={(1.03,0)}},
tick align=outside,
tick pos=right,
xtick pos=left,
title={SSIM},
x grid style={darkgray176},
xmin=1.5, xmax=6.5,
xtick={2,3,...,6},
xtick style={color=black},
y grid style={darkgray176},
ymin=0.784488011504296, ymax=0.981160706070834,
ytick style={color=black},
]
\addplot [black]
table {%
1.75 0.970322344908592
2.25 0.970322344908592
2.25 0.971114714821491
1.75 0.971114714821491
1.75 0.970322344908592
};
\addplot [black]
table {%
2 0.970322344908592
2 0.969633285702022
};
\addplot [black]
table {%
2 0.971114714821491
2 0.972221038135992
};
\addplot [black]
table {%
1.875 0.969633285702022
2.125 0.969633285702022
};
\addplot [black]
table {%
1.875 0.972221038135992
2.125 0.972221038135992
};
\addplot [black]
table {%
2.75 0.948256868206402
3.25 0.948256868206402
3.25 0.948913587375673
2.75 0.948913587375673
2.75 0.948256868206402
};
\addplot [black]
table {%
3 0.948256868206402
3 0.948193046829539
};
\addplot [black]
table {%
3 0.948913587375673
3 0.949895819555421
};
\addplot [black]
table {%
2.875 0.948193046829539
3.125 0.948193046829539
};
\addplot [black]
table {%
2.875 0.949895819555421
3.125 0.949895819555421
};
\addplot [black, mark=o, mark size=3, mark options={solid,fill opacity=0}, only marks]
table {%
3 0.94630368613635
};
\addplot [black]
table {%
3.75 0.906233515790045
4.25 0.906233515790045
4.25 0.906731863573939
3.75 0.906731863573939
3.75 0.906233515790045
};
\addplot [black]
table {%
4 0.906233515790045
4 0.906233515790045
};
\addplot [black]
table {%
4 0.906731863573939
4 0.906731863573939
};
\addplot [black]
table {%
3.875 0.906233515790045
4.125 0.906233515790045
};
\addplot [black]
table {%
3.875 0.906731863573939
4.125 0.906731863573939
};
\addplot [black, mark=o, mark size=3, mark options={solid,fill opacity=0}, only marks]
table {%
4 0.905075769504434
4 0.904739253320991
4 0.907984642681555
4 0.911040669039892
};
\addplot [black]
table {%
4.75 0.856917747509389
5.25 0.856917747509389
5.25 0.861753563696669
4.75 0.861753563696669
4.75 0.856917747509389
};
\addplot [black]
table {%
5 0.856917747509389
5 0.853278476729426
};
\addplot [black]
table {%
5 0.861753563696669
5 0.863129389130955
};
\addplot [black]
table {%
4.875 0.853278476729426
5.125 0.853278476729426
};
\addplot [black]
table {%
4.875 0.863129389130955
5.125 0.863129389130955
};
\addplot [black]
table {%
5.75 0.795183174976272
6.25 0.795183174976272
6.25 0.799258682099347
5.75 0.799258682099347
5.75 0.795183174976272
};
\addplot [black]
table {%
6 0.795183174976272
6 0.793427679439139
};
\addplot [black]
table {%
6 0.799258682099347
6 0.800943165403694
};
\addplot [black]
table {%
5.875 0.793427679439139
6.125 0.793427679439139
};
\addplot [black]
table {%
5.875 0.800943165403694
6.125 0.800943165403694
};
\addplot [darkorange25512714]
table {%
1.75 0.97049729171993
2.25 0.97049729171993
};
\addplot [darkorange25512714]
table {%
2.75 0.948510086374528
3.25 0.948510086374528
};
\addplot [darkorange25512714]
table {%
3.75 0.906598601461801
4.25 0.906598601461801
};
\addplot [darkorange25512714]
table {%
4.75 0.861486676429391
5.25 0.861486676429391
};
\addplot [darkorange25512714]
table {%
5.75 0.796399069469572
6.25 0.796399069469572
};
\end{groupplot}

\end{tikzpicture}

%% file: figs/fabian_nomotion_1_inverse_crime.tex
\def\imwidth{.185\textwidth}
\def\hs{1pt}
\def\vs{1pt}
\begin{tikzpicture}

\definecolor{darkgray176}{RGB}{176,176,176}

\begin{groupplot}[group style={group size=3 by 3, horizontal sep=\hs, vertical sep=\vs}, ticks=none, width=\imwidth,
title style={yshift=\titleshift},axis equal image]
\nextgroupplot[
tick align=outside,
title={a) LR
},
x grid style={darkgray176},
xmin=-0.5, xmax=95.5,
xtick pos=right,
xtick style={color=black},
xtick={-20,0,20,40,60,80,100},
xticklabels={0.0,0.5,1.0,,,,},
y dir=reverse,
ylabel={Axial},
ymin=-0.5, ymax=95.5,
ytick pos=left
]
\addplot graphics [includegraphics cmd=\pgfimage,xmin=-0.5, xmax=95.5, ymin=95.5, ymax=-0.5] {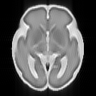};
\draw[red,very thin] (axis cs:34.1020691684181,24.1770057251406) rectangle (axis cs:60.3166962555751,49.2761167660356);
\nextgroupplot[
tick align=outside,
title={b) SAIR
},
x grid style={darkgray176},
xmin=-0.5, xmax=95.5,
xtick pos=right,
xtick style={color=black},
xtick={-20,0,20,40,60,80,100},
xticklabels={0.0,0.5,1.0,,,,},
y dir=reverse,
y grid style={darkgray176},
ymin=-0.5, ymax=95.5,
ytick pos=left,
ytick style={color=black},
ytick={-20,0,20,40,60,80,100},
yticklabels={0.0,0.5,1.0,,,,}
]
\addplot graphics [includegraphics cmd=\pgfimage,xmin=-0.5, xmax=95.5, ymin=95.5, ymax=-0.5] {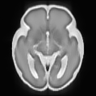};
\draw[red,very thin] (axis cs:34.1020691684181,24.1770057251406) rectangle (axis cs:60.3166962555751,49.2761167660356);

\nextgroupplot[
tick align=outside,
title={c) GT},
x grid style={darkgray176},
xmin=-0.5, xmax=95.5,
xtick pos=right,
xtick style={color=black},
xtick={-20,0,20,40,60,80,100},
xticklabels={0.0,0.5,1.0,,,,},
y dir=reverse,
y grid style={darkgray176},
ymin=-0.5, ymax=95.5,
ytick pos=left,
ytick style={color=black},
ytick={-20,0,20,40,60,80,100},
yticklabels={0.0,0.5,1.0,,,,}
]
\addplot graphics [includegraphics cmd=\pgfimage,xmin=-0.5, xmax=95.5, ymin=95.5, ymax=-0.5] {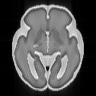};
\draw[red,very thin] (axis cs:34.1020691684181,24.1770057251406) rectangle (axis cs:60.3166962555751,49.2761167660356);

\nextgroupplot[
tick align=outside,
x grid style={darkgray176},
xmin=-0.5, xmax=95.5,
xtick pos=right,
xtick style={color=black},
xtick={-20,0,20,40,60,80,100},
xticklabels={0.0,0.5,1.0,,,,},
y dir=reverse,
ylabel={Sagittal},
ymin=-0.5, ymax=95.5,
ytick pos=left
]
\addplot graphics [includegraphics cmd=\pgfimage,xmin=-0.5, xmax=95.5, ymin=95.5, ymax=-0.5] {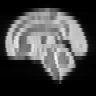};
\draw[red,very thin] (axis cs:34.1020691684181,24.1770057251406) rectangle (axis cs:60.3166962555751,49.2761167660356);

\nextgroupplot[
tick align=outside,
x grid style={darkgray176},
xmin=-0.5, xmax=95.5,
xtick pos=right,
xtick style={color=black},
xtick={-20,0,20,40,60,80,100},
xticklabels={0.0,0.5,1.0,,,,},
y dir=reverse,
y grid style={darkgray176},
ymin=-0.5, ymax=95.5,
ytick pos=left,
ytick style={color=black},
ytick={-20,0,20,40,60,80,100},
yticklabels={0.0,0.5,1.0,,,,}
]
\addplot graphics [includegraphics cmd=\pgfimage,xmin=-0.5, xmax=95.5, ymin=95.5, ymax=-0.5] {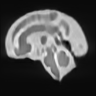};
\draw[red,very thin] (axis cs:34.1020691684181,24.1770057251406) rectangle (axis cs:60.3166962555751,49.2761167660356);

\nextgroupplot[
tick align=outside,
x grid style={darkgray176},
xmin=-0.5, xmax=95.5,
xtick pos=right,
xtick style={color=black},
xtick={-20,0,20,40,60,80,100},
xticklabels={0.0,0.5,1.0,,,,},
y dir=reverse,
y grid style={darkgray176},
ymin=-0.5, ymax=95.5,
ytick pos=left,
ytick style={color=black},
ytick={-20,0,20,40,60,80,100},
yticklabels={0.0,0.5,1.0,,,,}
]
\addplot graphics [includegraphics cmd=\pgfimage,xmin=-0.5, xmax=95.5, ymin=95.5, ymax=-0.5] {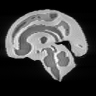};
\draw[red,very thin] (axis cs:34.1020691684181,24.1770057251406) rectangle (axis cs:60.3166962555751,49.2761167660356);

\nextgroupplot[
tick align=outside,
x grid style={darkgray176},
xmin=-0.5, xmax=95.5,
xtick pos=right,
xtick style={color=black},
xtick={-20,0,20,40,60,80,100},
xticklabels={0.0,0.5,1.0,,,,},
y dir=reverse,
ylabel={Coronal},
ymin=-0.5, ymax=95.5,
ytick pos=left
]
\addplot graphics [includegraphics cmd=\pgfimage,xmin=-0.5, xmax=95.5, ymin=95.5, ymax=-0.5] {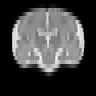};
\node at (axis cs:49.5,90) {\color{yellow} \tiny SSIM: 0.720};
\draw[red,very thin] (axis cs:34.1020691684181,24.1770057251406) rectangle (axis cs:60.3166962555751,49.2761167660356);

\nextgroupplot[
tick align=outside,
x grid style={darkgray176},
xmin=-0.5, xmax=95.5,
xtick pos=right,
xtick style={color=black},
xtick={-20,0,20,40,60,80,100},
xticklabels={0.0,0.5,1.0,,,,},
y dir=reverse,
y grid style={darkgray176},
ymin=-0.5, ymax=95.5,
ytick pos=left,
ytick style={color=black},
ytick={-20,0,20,40,60,80,100},
yticklabels={0.0,0.5,1.0,,,,}
]
\addplot graphics [includegraphics cmd=\pgfimage,xmin=-0.5, xmax=95.5, ymin=95.5, ymax=-0.5] {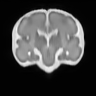};
\node at (axis cs:49.5,90) {\color{yellow} \tiny SSIM: 0.907};
\draw[red,very thin] (axis cs:34.1020691684181,24.1770057251406) rectangle (axis cs:60.3166962555751,49.2761167660356);

\nextgroupplot[
tick align=outside,
x grid style={darkgray176},
xmin=-0.5, xmax=95.5,
xtick pos=right,
xtick style={color=black},
xtick={-20,0,20,40,60,80,100},
xticklabels={0.0,0.5,1.0,,,,},
y dir=reverse,
y grid style={darkgray176},
ymin=-0.5, ymax=95.5,
ytick pos=left,
ytick style={color=black},
ytick={-20,0,20,40,60,80,100},
yticklabels={0.0,0.5,1.0,,,,}
]
\addplot graphics [includegraphics cmd=\pgfimage,xmin=-0.5, xmax=95.5, ymin=95.5, ymax=-0.5] {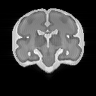};
\draw[red,very thin] (axis cs:34.1020691684181,24.1770057251406) rectangle (axis cs:60.3166962555751,49.2761167660356);
\end{groupplot}

\end{tikzpicture}
\begin{tikzpicture}

\definecolor{darkgray176}{RGB}{176,176,176}

\begin{groupplot}[group style={group size=3 by 3, horizontal sep=\hs, vertical sep=\vs}, ticks=none, width=\imwidth,
title style={yshift=\titleshift},axis equal image]
\nextgroupplot[
tick align=outside,
title={a) LR},
x grid style={darkgray176},
xmin=34.1020691684181, xmax=60.3166962555751,
xtick pos=right,
xtick style={color=black},
xtick={30,40,50,60,70},
xticklabels={0.0,0.5,1.0,,},
y dir=reverse,
ymin=24.1770057251406, ymax=49.2761167660356,
ytick pos=left
]
\addplot graphics [includegraphics cmd=\pgfimage,xmin=-0.5, xmax=95.5, ymin=95.5, ymax=-0.5] {figs/fabian_nomotion_1_inverse_crime/fabian_nomotion_1_inverse_crime-000.png};

\nextgroupplot[
tick align=outside,
title={b) SAIR},
x grid style={darkgray176},
xmin=34.1020691684181, xmax=60.3166962555751,
xtick pos=right,
xtick style={color=black},
xtick={30,40,50,60,70},
xticklabels={0.0,0.5,1.0,,},
y dir=reverse,
y grid style={darkgray176},
ymin=24.1770057251406, ymax=49.2761167660356,
ytick pos=left,
ytick style={color=black},
ytick={20,25,30,35,40,45,50},
yticklabels={0.0,0.5,1.0,,,,}
]
\addplot graphics [includegraphics cmd=\pgfimage,xmin=-0.5, xmax=95.5, ymin=95.5, ymax=-0.5] {figs/fabian_nomotion_1_inverse_crime/fabian_nomotion_1_inverse_crime-001.png};

\nextgroupplot[
tick align=outside,
title={c) GT},
x grid style={darkgray176},
xmin=34.1020691684181, xmax=60.3166962555751,
xtick pos=right,
xtick style={color=black},
xtick={30,40,50,60,70},
xticklabels={0.0,0.5,1.0,,},
y dir=reverse,
y grid style={darkgray176},
ymin=24.1770057251406, ymax=49.2761167660356,
ytick pos=left,
ytick style={color=black},
ytick={20,25,30,35,40,45,50},
yticklabels={0.0,0.5,1.0,,,,}
]
\addplot graphics [includegraphics cmd=\pgfimage,xmin=-0.5, xmax=95.5, ymin=95.5, ymax=-0.5] {figs/fabian_nomotion_1_inverse_crime/fabian_nomotion_1_inverse_crime-002.png};

\nextgroupplot[
tick align=outside,
x grid style={darkgray176},
xmin=34.1020691684181, xmax=60.3166962555751,
xtick pos=right,
xtick style={color=black},
xtick={30,40,50,60,70},
xticklabels={0.0,0.5,1.0,,},
y dir=reverse,
ymin=24.1770057251406, ymax=49.2761167660356,
ytick pos=left
]
\addplot graphics [includegraphics cmd=\pgfimage,xmin=-0.5, xmax=95.5, ymin=95.5, ymax=-0.5] {figs/fabian_nomotion_1_inverse_crime/fabian_nomotion_1_inverse_crime-003.png};

\nextgroupplot[
tick align=outside,
x grid style={darkgray176},
xmin=34.1020691684181, xmax=60.3166962555751,
xtick pos=right,
xtick style={color=black},
xtick={30,40,50,60,70},
xticklabels={0.0,0.5,1.0,,},
y dir=reverse,
y grid style={darkgray176},
ymin=24.1770057251406, ymax=49.2761167660356,
ytick pos=left,
ytick style={color=black},
ytick={20,25,30,35,40,45,50},
yticklabels={0.0,0.5,1.0,,,,}
]
\addplot graphics [includegraphics cmd=\pgfimage,xmin=-0.5, xmax=95.5, ymin=95.5, ymax=-0.5] {figs/fabian_nomotion_1_inverse_crime/fabian_nomotion_1_inverse_crime-004.png};

\nextgroupplot[
tick align=outside,
x grid style={darkgray176},
xmin=34.1020691684181, xmax=60.3166962555751,
xtick pos=right,
xtick style={color=black},
xtick={30,40,50,60,70},
xticklabels={0.0,0.5,1.0,,},
y dir=reverse,
y grid style={darkgray176},
ymin=24.1770057251406, ymax=49.2761167660356,
ytick pos=left,
ytick style={color=black},
ytick={20,25,30,35,40,45,50},
yticklabels={0.0,0.5,1.0,,,,}
]
\addplot graphics [includegraphics cmd=\pgfimage,xmin=-0.5, xmax=95.5, ymin=95.5, ymax=-0.5] {figs/fabian_nomotion_1_inverse_crime/fabian_nomotion_1_inverse_crime-005.png};

\nextgroupplot[
tick align=outside,
x grid style={darkgray176},
xmin=34.1020691684181, xmax=60.3166962555751,
xtick pos=right,
xtick style={color=black},
xtick={30,40,50,60,70},
xticklabels={0.0,0.5,1.0,,},
y dir=reverse,
ymin=24.1770057251406, ymax=49.2761167660356,
ytick pos=left
]
\addplot graphics [includegraphics cmd=\pgfimage,xmin=-0.5, xmax=95.5, ymin=95.5, ymax=-0.5] {figs/fabian_nomotion_1_inverse_crime/fabian_nomotion_1_inverse_crime-006.png};

\nextgroupplot[
tick align=outside,
x grid style={darkgray176},
xmin=34.1020691684181, xmax=60.3166962555751,
xtick pos=right,
xtick style={color=black},
xtick={30,40,50,60,70},
xticklabels={0.0,0.5,1.0,,},
y dir=reverse,
y grid style={darkgray176},
ymin=24.1770057251406, ymax=49.2761167660356,
ytick pos=left,
ytick style={color=black},
ytick={20,25,30,35,40,45,50},
yticklabels={0.0,0.5,1.0,,,,}
]
\addplot graphics [includegraphics cmd=\pgfimage,xmin=-0.5, xmax=95.5, ymin=95.5, ymax=-0.5] {figs/fabian_nomotion_1_inverse_crime/fabian_nomotion_1_inverse_crime-007.png};

\nextgroupplot[
tick align=outside,
x grid style={darkgray176},
xmin=34.1020691684181, xmax=60.3166962555751,
xtick pos=right,
xtick style={color=black},
xtick={30,40,50,60,70},
xticklabels={0.0,0.5,1.0,,},
y dir=reverse,
y grid style={darkgray176},
ymin=24.1770057251406, ymax=49.2761167660356,
ytick pos=left,
ytick style={color=black},
ytick={20,25,30,35,40,45,50},
yticklabels={0.0,0.5,1.0,,,,}
]
\addplot graphics [includegraphics cmd=\pgfimage,xmin=-0.5, xmax=95.5, ymin=95.5, ymax=-0.5] {figs/fabian_nomotion_1_inverse_crime/fabian_nomotion_1_inverse_crime-008.png};
\end{groupplot}

\end{tikzpicture}

%% file: secs/SAIR.tex
SAIR, our proposed pipeline, is based on a simplified model of the acquired LR volume 
$\lowresvol$ with respect to the true volume $\highresvol$. In particular, we assume that
\begin{equation} \label{eq:simplified_model}
    \lowresvol = \downsampling{z} \blur \highresvol + \noise\,,
\end{equation}
where $\downsampling{\mathrm{z}}$ is the linear operator that performs downsampling along the $\mathrm{z}$ axis 
(by convention, 
the axial direction of the series) by a factor of $r$, $\blur$ is a convolutional operator that 
approximates the frequency response of the MRI scanner (a.k.a. the slice-selection profile, here
chosen as in~\cite{art:Tourbier2015}), and $\noise$ is an additive white Gaussian noise. 

Building on \eqref{eq:simplified_model}, we propose to create a training set based on a simple upsampling 
$\upsampledvol = \upsampling{\mathrm{z}} \lowresvol$ (\emph{e.g.}, bicubic) of the data volume 
along $\mathrm{z}$ as the training target, and an artificially degraded volume using model 
\eqref{eq:simplified_model} along $\mathrm{x}$ as the training input $\downsampling{\mathrm{x}} \rotblur \upsampledvol + \noise$,
where $\rotblur$ is an axis-permuted version of $\blur$ (switching the roles of $\mathrm{z}$ and $\mathrm{x}$). If the model
represents the distortion caused by the MRI in $\mathrm{z}$ well enough, 
then a network that would perform well on this problem should be able to reconstruct a volume with isotropic resolution from the 
low-resolution volume $\lowresvol$, up to a permutation in the $\mathrm{x}$ and $\mathrm{z}$ axes. Because there is nothing special about
the $\mathrm{x}$ axis, any rotation by an angle $\theta$ around the $\mathrm{z}$ axis of the upsampled volume 
$\rotation \upsampledvol$ is equally valid to augment the training dataset. Thus, we create the 
training dataset from $n_{\mathrm{train}}$ rotated volumes 
for evenly spaced angles between \unit{0}{\degree} and \unit{180}{\degree}.
This pipeline is summarized in Figure~\ref{fig:SAIR_train}. There, 
we see that the training inputs are upsampled in the $\mathrm{x}$ axis, so that the network is only 
trained to refine the upsampling operator of choice, $\upsampling{\mathrm{x}}$.
All the results in this paper were obtained with 
$n_\mathrm{train}=10$ and $\upsampling{\mathrm{z}}$ the bicubic upsampling operator.

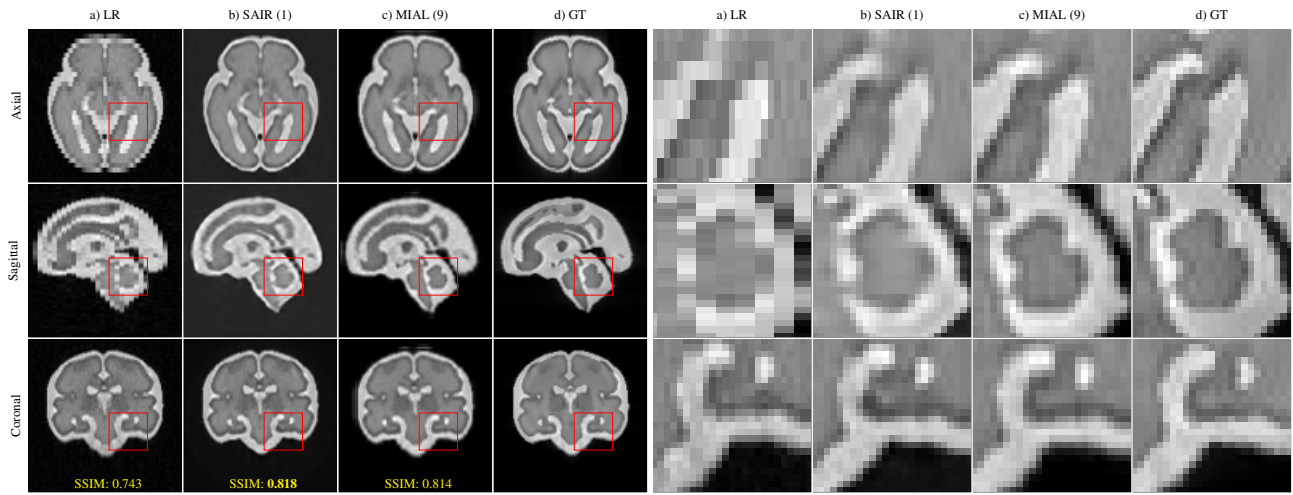
\begin{figure*}
    \centering
    \input{figs/fabian_nomotion_4.tex}

    \vspace{-5pt}
    
    \caption{Left panel: a) synthetic LR HASTE series simulated in the coronal orientation using FaBiAN for a subject of 30 weeks of gestational age; b) SAIR reconstruction from that single LR series; c) MIALSRTK reconstruction combining nine orthogonal LR series (MIAL). The SSIM was computed with respect to d) a synthetic 3D isotropic high resolution ground-truth image.
    Right panel: Zoom over regions of interest. 
    \label{fig:fabian_results}}
\end{figure*}

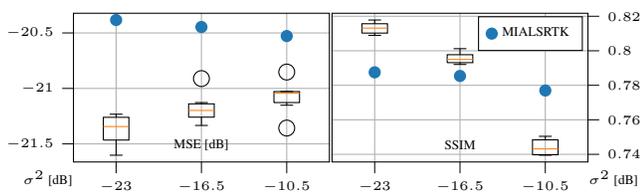
\begin{figure}
    \centering
    \input{figs/ISBI_SNR_study}
    
    \vspace{-18pt}
    
    \caption{Reconstruction performance as a function of the noise level.
        Reported for a single reconstruction with MIALSRTK using nine volumes, and in the 
        form of a boxplot for SAIR for six different noise realizations. 
    \label{fig:SNR_study}}
\end{figure}
            
At prediction time, we take an ensemble approach, combining the predictions of the network when applied to 
$n_{\mathrm{pred}}$ different rotations of the upsampled volume between \unit{0}{\degree} and 
\unit{180}{\degree} (see Figure~\ref{fig:SAIR_pred}). To combine these predictions, we use the Fourier-burst accumulation 
technique~\cite{FBA}, summarized by
\begin{equation}
    \predvol = \mathrm{FFT}^{-1}\left\lbrace \sum_{m=1}^{n_{\mathrm{pred}}} 
        \mathbf{W}_m \odot \hat{\mathbf{X}}_m^{\mathrm{pred}} \right\rbrace,
\end{equation}
where $\hat{\mathbf{X}}_m^{\mathrm{pred}}$ is the FFT of 
$\mathbf{X}_m^{\mathrm{pred}} = \rotationang{-\theta_m} \mathrm{NN}(\rotationang{\theta_m} \upsampledvol)$,
where $\mathrm{NN}(\cdot)$ represents the trained neural network. 
The weighting mask $\mathbf{W}_m$ is computed as
\begin{equation}
    \mathbf{W}_m = \frac{
            \left\vert \hat{\mathbf{X}}_m^{\mathrm{pred}}\right\vert^2}{
            \sum_{k=1}^{n_{\mathrm{pred}}} \left\vert \hat{\mathbf{X}}_k^{\mathrm{pred}} \right\vert^2},
\end{equation}
which corresponds to $p=2$ in the expressions in~\cite{FBA}.
All results in this paper were obtained with 
$n_\mathrm{pred}=15$.

SAIR is a proof-of-concept pipeline in the context of FBMRI. 
Model~\eqref{eq:simplified_model} captures the most dominant effects of MR acquisition, and 
Fourier-burst accumulation~\cite{FBA} combines frequency information from several independent
predictions. Our proposed architecture is a
slice-wise (2D) U-Net with $32$ convolutional initial channels, $(7 \times 7)$ convolutional kernels,
a skip connection, and a single encoder/decoder step. Due to the use of a U-Net with skip connection
in the fashion of~\cite{Jin2017}, the network is trained to correct 
the artifacts in the input image. The architecture is very shallow, which is an advantage in the 
self-supervised context, in which the network is to be trained for each volume. 

%% file: figs/fabian_nomotion_4.tex
\def\imwidth{.235\textwidth}
\def\hs{1pt}
\def\vs{1pt}
\begin{tikzpicture}

\definecolor{darkgray176}{RGB}{176,176,176}

\begin{groupplot}[group style={group size=4 by 3, horizontal sep=\hs, vertical sep=\vs}, ticks=none, width=\imwidth,
title style={yshift=\titleshift},axis equal image]

\nextgroupplot[
tick align=outside,
title={a) LR 
},
x grid style={darkgray176},
xmin=-0.5, xmax=95.5,
xtick pos=left,
xtick style={color=black},
y dir=reverse,
ylabel={{Axial}},
ymin=-0.5, ymax=95.5,
ytick pos=left
]
\addplot graphics [includegraphics cmd=\pgfimage,xmin=-0.5, xmax=95.5, ymin=95.5, ymax=-0.5] {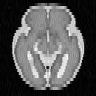};
\draw[red,very thin] (axis cs:49.9492803482519,45.7523665754306) rectangle (axis cs:74.1187940836723,69.1782029651457);

\nextgroupplot[
tick align=outside,
title={{b) SAIR (1)
}},
x grid style={darkgray176},
xmin=-0.5, xmax=95.5,
xtick pos=left,
xtick style={color=black},
y dir=reverse,
y grid style={darkgray176},
ymin=-0.5, ymax=95.5,
ytick pos=left,
ytick style={color=black}
]
\addplot graphics [includegraphics cmd=\pgfimage,xmin=-0.5, xmax=95.5, ymin=95.5, ymax=-0.5] {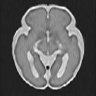};
\draw[red,very thin] (axis cs:49.9492803482519,45.7523665754306) rectangle (axis cs:74.1187940836723,69.1782029651457);

\nextgroupplot[
tick align=outside,
title={{c) MIAL (9) 
}},
x grid style={darkgray176},
xmin=-0.5, xmax=95.5,
xtick pos=left,
xtick style={color=black},
y dir=reverse,
y grid style={darkgray176},
ymin=-0.5, ymax=95.5,
ytick pos=left,
ytick style={color=black}
]
\addplot graphics [includegraphics cmd=\pgfimage,xmin=-0.5, xmax=95.5, ymin=95.5, ymax=-0.5] {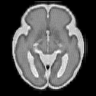};
\draw[red,very thin] (axis cs:49.9492803482519,45.7523665754306) rectangle (axis cs:74.1187940836723,69.1782029651457);

\nextgroupplot[
tick align=outside,
title={{d) GT}
},
x grid style={darkgray176},
xmin=-0.5, xmax=95.5,
xtick pos=left,
xtick style={color=black},
y dir=reverse,
y grid style={darkgray176},
ymin=-0.5, ymax=95.5,
ytick pos=left,
ytick style={color=black}
]
\addplot graphics [includegraphics cmd=\pgfimage,xmin=-0.5, xmax=95.5, ymin=95.5, ymax=-0.5] {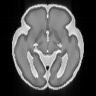};
\draw[red,very thin] (axis cs:49.9492803482519,45.7523665754306) rectangle (axis cs:74.1187940836723,69.1782029651457);

\nextgroupplot[
tick align=outside,
x grid style={darkgray176},
xmin=-0.5, xmax=95.5,
xtick pos=left,
xtick style={color=black},
y dir=reverse,
ylabel={{Sagittal}},
ymin=-0.5, ymax=95.5,
ytick pos=left
]
\addplot graphics [includegraphics cmd=\pgfimage,xmin=-0.5, xmax=95.5, ymin=95.5, ymax=-0.5] {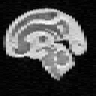};
\draw[red,very thin] (axis cs:49.9492803482519,45.7523665754306) rectangle (axis cs:74.1187940836723,69.1782029651457);

\nextgroupplot[
tick align=outside,
x grid style={darkgray176},
xmin=-0.5, xmax=95.5,
xtick pos=left,
xtick style={color=black},
y dir=reverse,
y grid style={darkgray176},
ymin=-0.5, ymax=95.5,
ytick pos=left,
ytick style={color=black}
]
\addplot graphics [includegraphics cmd=\pgfimage,xmin=-0.5, xmax=95.5, ymin=95.5, ymax=-0.5] {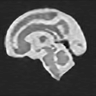};
\draw[red,very thin] (axis cs:49.9492803482519,45.7523665754306) rectangle (axis cs:74.1187940836723,69.1782029651457);

\nextgroupplot[
tick align=outside,
x grid style={darkgray176},
xmin=-0.5, xmax=95.5,
xtick pos=left,
xtick style={color=black},
y dir=reverse,
y grid style={darkgray176},
ymin=-0.5, ymax=95.5,
ytick pos=left,
ytick style={color=black}
]
\addplot graphics [includegraphics cmd=\pgfimage,xmin=-0.5, xmax=95.5, ymin=95.5, ymax=-0.5] {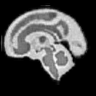};
\draw[red,very thin] (axis cs:49.9492803482519,45.7523665754306) rectangle (axis cs:74.1187940836723,69.1782029651457);

\nextgroupplot[
tick align=outside,
x grid style={darkgray176},
xmin=-0.5, xmax=95.5,
xtick pos=left,
xtick style={color=black},
y dir=reverse,
y grid style={darkgray176},
ymin=-0.5, ymax=95.5,
ytick pos=left,
ytick style={color=black}
]
\addplot graphics [includegraphics cmd=\pgfimage,xmin=-0.5, xmax=95.5, ymin=95.5, ymax=-0.5] {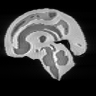};
\draw[red,very thin] (axis cs:49.9492803482519,45.7523665754306) rectangle (axis cs:74.1187940836723,69.1782029651457);

\nextgroupplot[
tick align=outside,
x grid style={darkgray176},
xmin=-0.5, xmax=95.5,
xtick pos=left,
xtick style={color=black},
y dir=reverse,
ylabel={{Coronal}},
ymin=-0.5, ymax=95.5,
ytick pos=left
]
\addplot graphics [includegraphics cmd=\pgfimage,xmin=-0.5, xmax=95.5, ymin=95.5, ymax=-0.5] {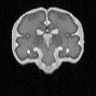};
\node at (axis cs:49.5,90) {\color{yellow} \tiny 
SSIM: 0.743};
\draw[red,very thin] (axis cs:49.9492803482519,45.7523665754306) rectangle (axis cs:74.1187940836723,69.1782029651457);

\nextgroupplot[
tick align=outside,
x grid style={darkgray176},
xmin=-0.5, xmax=95.5,
xtick pos=left,
xtick style={color=black},
y dir=reverse,
y grid style={darkgray176},
ymin=-0.5, ymax=95.5,
ytick pos=left,
ytick style={color=black}
]
\addplot graphics [includegraphics cmd=\pgfimage,xmin=-0.5, xmax=95.5, ymin=95.5, ymax=-0.5] {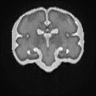};
\node at (axis cs:49.5,90) {\color{yellow} \tiny 
SSIM: \textbf{0.818}};
\draw[red,very thin] (axis cs:49.9492803482519,45.7523665754306) rectangle (axis cs:74.1187940836723,69.1782029651457);

\nextgroupplot[
tick align=outside,
x grid style={darkgray176},
xmin=-0.5, xmax=95.5,
xtick pos=left,
xtick style={color=black},
y dir=reverse,
y grid style={darkgray176},
ymin=-0.5, ymax=95.5,
ytick pos=left,
ytick style={color=black}
]
\addplot graphics [includegraphics cmd=\pgfimage,xmin=-0.5, xmax=95.5, ymin=95.5, ymax=-0.5] {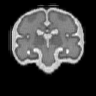};
\node at (axis cs:49.5,90) {\color{yellow} \tiny 
SSIM: 0.814};
\draw[red,very thin] (axis cs:49.9492803482519,45.7523665754306) rectangle (axis cs:74.1187940836723,69.1782029651457);

\nextgroupplot[
tick align=outside,
x grid style={darkgray176},
xmin=-0.5, xmax=95.5,
xtick pos=left,
xtick style={color=black},
y dir=reverse,
y grid style={darkgray176},
ymin=-0.5, ymax=95.5,
ytick pos=left,
ytick style={color=black}
]
\addplot graphics [includegraphics cmd=\pgfimage,xmin=-0.5, xmax=95.5, ymin=95.5, ymax=-0.5] {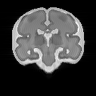};
\draw[red,very thin] (axis cs:49.9492803482519,45.7523665754306) rectangle (axis cs:74.1187940836723,69.1782029651457);
\end{groupplot}
\end{tikzpicture}
\begin{tikzpicture}
\definecolor{darkgray176}{RGB}{176,176,176}
\begin{groupplot}[group style={group size=4 by 3, horizontal sep=\hs, vertical sep=\vs}, ticks=none, width=\imwidth,
title style={yshift=\titleshift},axis equal image]
\nextgroupplot[
tick align=outside,
title={{a) LR
}},
x grid style={darkgray176},
xmin=49.9492803482519, xmax=74.1187940836723,
xtick pos=left,
xtick style={color=black},
y dir=reverse,
ymin=45.7523665754306, ymax=69.1782029651457,
ytick pos=left
]
\addplot graphics [includegraphics cmd=\pgfimage,xmin=-0.5, xmax=95.5, ymin=95.5, ymax=-0.5] {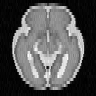};

\nextgroupplot[
tick align=outside,
title={{b) SAIR (1)
}},
x grid style={darkgray176},
xmin=49.9492803482519, xmax=74.1187940836723,
xtick pos=left,
xtick style={color=black},
y dir=reverse,
y grid style={darkgray176},
ymin=45.7523665754306, ymax=69.1782029651457,
ytick pos=left,
ytick style={color=black}
]
\addplot graphics [includegraphics cmd=\pgfimage,xmin=-0.5, xmax=95.5, ymin=95.5, ymax=-0.5] {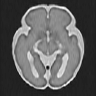};

\nextgroupplot[
tick align=outside,
title={{c) MIAL (9)
}},
x grid style={darkgray176},
xmin=49.9492803482519, xmax=74.1187940836723,
xtick pos=left,
xtick style={color=black},
y dir=reverse,
y grid style={darkgray176},
ymin=45.7523665754306, ymax=69.1782029651457,
ytick pos=left,
ytick style={color=black}
]
\addplot graphics [includegraphics cmd=\pgfimage,xmin=-0.5, xmax=95.5, ymin=95.5, ymax=-0.5] {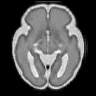};

\nextgroupplot[
tick align=outside,
title={{d) GT}
},
x grid style={darkgray176},
xmin=49.9492803482519, xmax=74.1187940836723,
xtick pos=left,
xtick style={color=black},
y dir=reverse,
y grid style={darkgray176},
ymin=45.7523665754306, ymax=69.1782029651457,
ytick pos=left,
ytick style={color=black}
]
\addplot graphics [includegraphics cmd=\pgfimage,xmin=-0.5, xmax=95.5, ymin=95.5, ymax=-0.5] {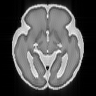};

\nextgroupplot[
tick align=outside,
x grid style={darkgray176},
xmin=49.9492803482519, xmax=74.1187940836723,
xtick pos=left,
xtick style={color=black},
y dir=reverse,
ymin=45.7523665754306, ymax=69.1782029651457,
ytick pos=left
]
\addplot graphics [includegraphics cmd=\pgfimage,xmin=-0.5, xmax=95.5, ymin=95.5, ymax=-0.5] {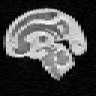};

\nextgroupplot[
tick align=outside,
x grid style={darkgray176},
xmin=49.9492803482519, xmax=74.1187940836723,
xtick pos=left,
xtick style={color=black},
y dir=reverse,
y grid style={darkgray176},
ymin=45.7523665754306, ymax=69.1782029651457,
ytick pos=left,
ytick style={color=black}
]
\addplot graphics [includegraphics cmd=\pgfimage,xmin=-0.5, xmax=95.5, ymin=95.5, ymax=-0.5] {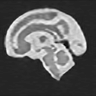};

\nextgroupplot[
tick align=outside,
x grid style={darkgray176},
xmin=49.9492803482519, xmax=74.1187940836723,
xtick pos=left,
xtick style={color=black},
y dir=reverse,
y grid style={darkgray176},
ymin=45.7523665754306, ymax=69.1782029651457,
ytick pos=left,
ytick style={color=black}
]
\addplot graphics [includegraphics cmd=\pgfimage,xmin=-0.5, xmax=95.5, ymin=95.5, ymax=-0.5] {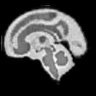};

\nextgroupplot[
tick align=outside,
x grid style={darkgray176},
xmin=49.9492803482519, xmax=74.1187940836723,
xtick pos=left,
xtick style={color=black},
y dir=reverse,
y grid style={darkgray176},
ymin=45.7523665754306, ymax=69.1782029651457,
ytick pos=left,
ytick style={color=black}
]
\addplot graphics [includegraphics cmd=\pgfimage,xmin=-0.5, xmax=95.5, ymin=95.5, ymax=-0.5] {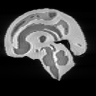};

\nextgroupplot[
tick align=outside,
x grid style={darkgray176},
xmin=49.9492803482519, xmax=74.1187940836723,
xtick pos=left,
xtick style={color=black},
y dir=reverse,
ymin=45.7523665754306, ymax=69.1782029651457,
ytick pos=left
]
\addplot graphics [includegraphics cmd=\pgfimage,xmin=-0.5, xmax=95.5, ymin=95.5, ymax=-0.5] {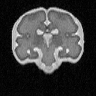};

\nextgroupplot[
tick align=outside,
x grid style={darkgray176},
xmin=49.9492803482519, xmax=74.1187940836723,
xtick pos=left,
xtick style={color=black},
y dir=reverse,
y grid style={darkgray176},
ymin=45.7523665754306, ymax=69.1782029651457,
ytick pos=left,
ytick style={color=black}
]
\addplot graphics [includegraphics cmd=\pgfimage,xmin=-0.5, xmax=95.5, ymin=95.5, ymax=-0.5] {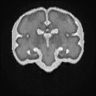};

\nextgroupplot[
tick align=outside,
x grid style={darkgray176},
xmin=49.9492803482519, xmax=74.1187940836723,
xtick pos=left,
xtick style={color=black},
y dir=reverse,
y grid style={darkgray176},
ymin=45.7523665754306, ymax=69.1782029651457,
ytick pos=left,
ytick style={color=black}
]
\addplot graphics [includegraphics cmd=\pgfimage,xmin=-0.5, xmax=95.5, ymin=95.5, ymax=-0.5] {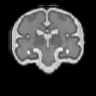};

\nextgroupplot[
tick align=outside,
x grid style={darkgray176},
xmin=49.9492803482519, xmax=74.1187940836723,
xtick pos=left,
xtick style={color=black},
y dir=reverse,
y grid style={darkgray176},
ymin=45.7523665754306, ymax=69.1782029651457,
ytick pos=left,
ytick style={color=black}
]
\addplot graphics [includegraphics cmd=\pgfimage,xmin=-0.5, xmax=95.5, ymin=95.5, ymax=-0.5] {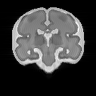};
\end{groupplot}

\end{tikzpicture}

%% file: figs/ISBI_SNR_study.tex
\def\hs{1pt}
\def\vs{1pt}
\begin{tikzpicture}
\definecolor{darkgray176}{RGB}{176,176,176}
\definecolor{darkorange25512714}{RGB}{255,127,14}
\definecolor{steelblue31119180}{RGB}{31,119,180}
\begin{groupplot}[group style={group size=2 by 1,horizontal sep=\hs, vertical sep=\vs},height={.2\textwidth}, width={.28\textwidth},grid=both,title style={anchor=center,at={(0.5,0.01)}},]
\nextgroupplot[
tick align=outside,
tick pos=left,
title={MSE [dB]},
x grid style={darkgray176},
xmin=0.5, xmax=3.5,
xtick style={color=black},
y grid style={darkgray176},
ymin=-21.6628326928724, ymax=-20.3226369239407,
ytick style={color=black},
xtick={1,2,3},
xticklabels={$-23$, $-16.5$, $-10.5$},
xlabel={$\sigma^2$~[dB]},
x label style={at={(-0.1,0)}},
]
\addplot [black, forget plot]
table {%
0.85 -21.4654229536495
1.15 -21.4654229536495
1.15 -21.2602692071439
0.85 -21.2602692071439
0.85 -21.4654229536495
};
\addplot [black, forget plot]
table {%
1 -21.4654229536495
1 -21.6019147033755
};
\addplot [black, forget plot]
table {%
1 -21.2602692071439
1 -21.2323901909704
};
\addplot [black, forget plot]
table {%
0.925 -21.6019147033755
1.075 -21.6019147033755
};
\addplot [black, forget plot]
table {%
0.925 -21.2323901909704
1.075 -21.2323901909704
};
\addplot [black, forget plot]
table {%
1.85 -21.2589879918481
2.15 -21.2589879918481
2.15 -21.1390929613624
1.85 -21.1390929613624
1.85 -21.2589879918481
};
\addplot [black, forget plot]
table {%
2 -21.2589879918481
2 -21.3336523636285
};
\addplot [black, forget plot]
table {%
2 -21.1390929613624
2 -21.1275482680527
};
\addplot [black, forget plot]
table {%
1.925 -21.3336523636285
2.075 -21.3336523636285
};
\addplot [black, forget plot]
table {%
1.925 -21.1275482680527
2.075 -21.1275482680527
};
\addplot [black, mark=o, mark size=3, mark options={solid,fill opacity=0}, only marks, forget plot]
table {%
2 -20.911394883552
};
\addplot [black, forget plot]
table {%
2.85 -21.1274515816458
3.15 -21.1274515816458
3.15 -21.0283015833348
2.85 -21.0283015833348
2.85 -21.1274515816458
};
\addplot [black, forget plot]
table {%
3 -21.1274515816458
3 -21.1511286638523
};
\addplot [black, forget plot]
table {%
3 -21.0283015833348
3 -21.0276763063455
};
\addplot [black, forget plot]
table {%
2.925 -21.1511286638523
3.075 -21.1511286638523
};
\addplot [black, forget plot]
table {%
2.925 -21.0276763063455
3.075 -21.0276763063455
};
\addplot [black, mark=o, mark size=3, mark options={solid,fill opacity=0}, only marks, forget plot]
table {%
3 -21.3573137431321
3 -20.8523706927811
};
\addplot [semithick, steelblue31119180, only marks]
table {%
1 -20.3835549134376
2 -20.4465127507582
3 -20.5281478486029
};
\addplot [darkorange25512714, forget plot]
table {%
0.85 -21.3434121141401
1.15 -21.3434121141401
};
\addplot [darkorange25512714, forget plot]
table {%
1.85 -21.1985366045914
2.15 -21.1985366045914
};
\addplot [darkorange25512714, forget plot]
table {%
2.85 -21.0432988746646
3.15 -21.0432988746646
};

\nextgroupplot[
tick align=outside,
tick pos=left,
ytick pos=right,
title={SSIM},
x grid style={darkgray176},
xmin=0.5, xmax=3.5,
xtick style={color=black},
y grid style={darkgray176},
ymin=0.735577583714133, ymax=0.821738724737902,
ytick style={color=black},
legend entries={{\tiny MIALSRTK}},
xtick={1,2,3},
xticklabels={$-23$, $-16.5$, $-10.5$},
xlabel={$\sigma^2$~[dB]},
x label style={at={(1.1,0)}},
]
\addplot [black, forget plot]
table {%
0.85 0.810257792223784
1.15 0.810257792223784
1.15 0.815667564076219
0.85 0.815667564076219
0.85 0.810257792223784
};
\addplot [black, forget plot]
table {%
1 0.810257792223784
1 0.808948088943085
};
\addplot [black, forget plot]
table {%
1 0.815667564076219
1 0.817822309236821
};
\addplot [black, forget plot]
table {%
0.925 0.808948088943085
1.075 0.808948088943085
};
\addplot [black, forget plot]
table {%
0.925 0.817822309236821
1.075 0.817822309236821
};
\addplot [black, forget plot]
table {%
1.85 0.793125874237734
2.15 0.793125874237734
2.15 0.79772727839624
1.85 0.79772727839624
1.85 0.793125874237734
};
\addplot [black, forget plot]
table {%
2 0.793125874237734
2 0.792086714905061
};
\addplot [black, forget plot]
table {%
2 0.79772727839624
2 0.801224796778881
};
\addplot [black, forget plot]
table {%
1.925 0.792086714905061
2.075 0.792086714905061
};
\addplot [black, forget plot]
table {%
1.925 0.801224796778881
2.075 0.801224796778881
};
\addplot [black, forget plot]
table {%
2.85 0.739748598078315
3.15 0.739748598078315
3.15 0.748381543906726
2.85 0.748381543906726
2.85 0.739748598078315
};
\addplot [black, forget plot]
table {%
3 0.739748598078315
3 0.739493999215213
};
\addplot [black, forget plot]
table {%
3 0.748381543906726
3 0.750443730503112
};
\addplot [black, forget plot]
table {%
2.925 0.739493999215213
3.075 0.739493999215213
};
\addplot [black, forget plot]
table {%
2.925 0.750443730503112
3.075 0.750443730503112
};
\addplot [semithick, steelblue31119180, only marks]
table {%
1 0.787518664212104
2 0.785406110363533
3 0.776982516344007
};
\addplot [darkorange25512714, forget plot]
table {%
0.85 0.813147688965982
1.15 0.813147688965982
};
\addplot [darkorange25512714, forget plot]
table {%
1.85 0.794971894853004
2.15 0.794971894853004
};
\addplot [darkorange25512714, forget plot]
table {%
2.85 0.743138601838185
3.15 0.743138601838185
};
\end{groupplot}

\end{tikzpicture}

%% file: secs/data.tex
\subsection{Synthetic Low-Resolution Series}
    \label{sec:fabian_data}
    
        Synthetic, yet realistic T2w MR images of the fetal brain at different gestational ages (GA) were derived from a normative spatiotemporal MRI atlas of the fetal brain~\cite{Gholipour2017} through a simulated 
        fetal-brain MR acquisition of a numerical phantom (FaBiAN)~\cite{Lajous2022}. In this highly flexible and controlled environment, we have reproduced the clinical MR protocol for the half-Fourier acquisition single-shot turbo spin echo sequence (HASTE, Siemens Healthineers) performed at our local hospital (at \unit{1.5}{\tesla}: TR/TE, \unit{1200}{\milli\second}/\unit{90}{\milli\second}; flip angle, \unit{90}{\degree}; echo train length, $224$; echo spacing, \unit{4.08}{\milli\second}; field-of-view, $\unit{(360 \times 360)}{\squaren\mm}$; voxel size, $\unit{(1.1 \times 1.1 \times 3.0)}{\mm\cubed}$; inter-slice gap, $10\%$)~\cite{lajous_dataset_2020,khawam_fetal_2021}.

        For a subject of $30$ weeks of GA, nine orthogonal series (three in each orientation, with a shift of the field-of-view of \unit{1.6}{\mm} between each series) were simulated without motion. Random isotropic complex Gaussian noise (with mean 0 and standard deviation $\sigma_0=0.15$) was added to the k-space data to qualitatively match the noise characteristics of clinical acquisitions.
        Corresponding brain masks were automatically generated along with the 2D LR series.
        Additional data were simulated with two extra noise levels ($\sigma=0.07$ and $0.30$). Six independent realizations of the same LR series were generated for each noise level.

        A 3D HR isotropic HASTE image of the fetal brain was simulated for each subject without noise or motion to serve as a reference for the quantitative evaluation of the corresponding SR reconstructions. For comparison, a multiple-volume SR reconstruction was generated with MIALSRTK~\cite{art:Tourbier2015,tourbier_medical-image-analysis-laboratorymialsuperresolutiontoolkit_2020} using the nine simulated orthogonal LR series. 

    \subsection{Clinical Data}
        \label{sec:clinical_data}
    
        Clinical MR images at 1.5 T (MAGNETOM Aera, Siemens Healthcare, Erlangen, Germany) were retrospectively 
        selected for two normally-developing subjects of 28 and 33 weeks of GA. For comparison purposes, both 
        subjects were SR-reconstructed with the MIALSRTK pipeline~\cite{art:Tourbier2015,tourbier_medical-image-analysis-laboratorymialsuperresolutiontoolkit_2020} using all the LR series available for each fetus (three without
        relevant motion for the younger fetus and eight with little-to-moderate motion for the older fetus).
        The voxel size was $\unit{(1.125 \times 1.125 \times 3.3)}{\mm\cubed}$.

%% file: secs/empirical_results.tex
    \subsection{Results on Simulated Data}
        
        We first analyze the performance of SAIR quantitatively on synthetic data. Because the ground truth (GT) 
        HR images are available in this scenario, we evaluate reconstructions by computing the mean-squared error 
        (MSE) and the mean structural similarity index (SSIM) over the brain area of every volume, as indicated by
        the corresponding GT mask. 
        
        \subsubsection{Simplified MRI Model}
        \label{sec:results_simplified-model}
            
            We start by exploring the performance of SAIR with respect to the resolution ratio
            $r=\Delta_\mathrm{z}/\Delta_\mathrm{x}=\Delta_\mathrm{z}/\Delta_\mathrm{y}$. For this experiment, we use the simplified model
            \eqref{eq:simplified_model} to simulate the MRI acquisitions of the subject described
            in Section~\ref{sec:fabian_data}. In Figure~\ref{fig:r4-example}, we see an example of 
            the resulting LR volume, its reconstruction by SAIR, and the GT HR volume for $r=4$. 
            Given the coarse resolution of the original data, SAIR does an impressive 
            job at recovering many of the lost details and at improving the SSIM (MSE)
            from $0.720$ ($-19.05$~dB, respectively) in the LR volume to $0.907$ ($-22.57$~dB, respectively) in the reconstructed 
            volume. In Figure~\ref{fig:different_rs}, we report the performance of SAIR as a function 
            of $r$. As expected, it decreases when $r$ increases, but relatively high SSIMs 
            ($\approx 0.8$) are still achieved for the extreme case $r=6$.
            
        \subsubsection{Realistic Synthetic Data}
        
            We proceed by evaluating the performance of SAIR on the synthetic T2w MR images described in Section~\ref{sec:fabian_data}, compared to that of the MIALSRTK multiple-volume reconstruction
            with $9$ different volumes. In Figure~\ref{fig:fabian_results},
            a reconstruction obtained from SAIR (which uses a single volume) is compared to a reconstruction obtained 
            from MIALSRTK (which uses $9$ different volumes). There, we observe that 1) the reconstructions
            are of similar perceptual quality, with SAIR seeming advantageous in the overall view and MIALSRTK
            dominating in the zoomed view, and that 2) the obtained SSIMs are similar ($0.818$ for SAIR and 
            $0.814$ for MIALSRTK). Additionally, the MSEs are $-22.35$~dB for SAIR and $-21.18$~dB for MIALSRTK. 
            However, we suggest here that MSE is not a reliable metric to evaluate the quality of these reconstructions, as it does not seem to match the perceived quality.
            
            We further analyzed both how SAIR and MIALSRTK were affected by the level of noise. 
            In Figure~\ref{fig:SNR_study}, we show the performance of both reconstruction methods for the three
            noise levels specified in Section~\ref{sec:fabian_data}. We observe that SAIR is more sensitive 
            to noise. This reduced robustness was to be expected because SAIR uses only one volume, as opposed
            to the $9$ volumes used by MIALSRTK. Indeed, the trend in the evolution of the SSIM between MIALSRTK
            and SAIR is similar, with the slopes being much more pronounced in the case of SAIR. We 
            verify here that MSE is not a reliable metric to evaluate the quality of the reconstruction, as shown 
            by the discrepancy between MSE and SSIM changes for MIALSRTK.

        \subsection{Results on Clinical Data}
    
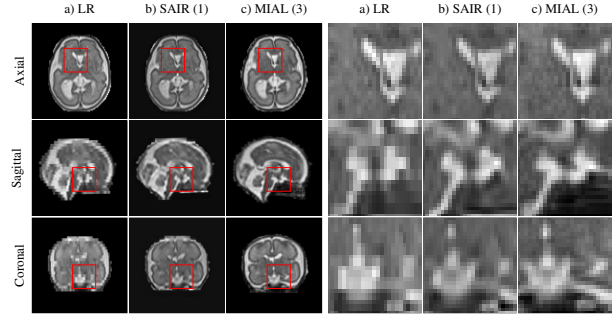
\begin{figure}
                \centering
                \input{figs/anat_nomotion_1} 
                \caption{Reconstruction of a clinical case without motion.\label{fig:clinical_results_1}}
            \end{figure}

\begin{figure}
                \centering
                \input{figs/anat_27_cor} 
                \caption{Reconstruction of a clinical case with little motion artefacts. \label{fig:clinical_results_2}}
            \end{figure}
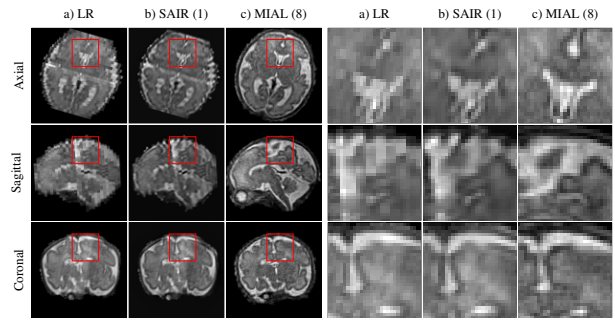

            Finally, we present the SR reconstructions using SAIR and MIALSRTK on two real clinical LR volumes, one without motion (Figure~\ref{fig:clinical_results_1}) and the other one with little amplitude of fetal movements (Figure~\ref{fig:clinical_results_2}). Overall, both methods provide very similar results. Although the reconstruction from one single volume with SAIR looks a little blurrier than the one combining multiple volumes with MIALSRTK, the quality of the reconstruction is remarkable, especially when considering that, as opposed to 
            MIALSRTK, SAIR does not include any mechanism to compensate for inter-slice motion.


%% file: figs/anat_nomotion_1.tex
\def\imwidth{.185\textwidth}\def\hs{1pt}\def\vs{1pt}
\begin{tikzpicture}

\definecolor{darkgray176}{RGB}{176,176,176}

\begin{groupplot}[group style={group size=3 by 3,horizontal sep=\hs, vertical sep=\vs}, ticks=none, width=\imwidth,
title style={yshift=\titleshift},axis equal image]
\nextgroupplot[
ylabel={Axial},
tick align=outside,
title={a) LR},
x grid style={darkgray176},
xmin=-0.5, xmax=99.5,
xtick pos=right,
xtick style={color=black},
xtick={-20,0,20,40,60,80,100},
xticklabels={0.0,0.5,1.0,,,,},
y dir=reverse,
ymin=-0.5, ymax=99.5,
ytick pos=left
]
\addplot graphics [includegraphics cmd=\pgfimage,xmin=-0.5, xmax=99.5, ymin=99.5, ymax=-0.5] {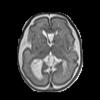};
\draw[red,very thin] (axis cs:33.08,25.63) rectangle (axis cs:57.48,50.62);

\nextgroupplot[
tick align=outside,
title={b) SAIR (1)},
x grid style={darkgray176},
xmin=-0.5, xmax=99.5,
xtick pos=right,
xtick style={color=black},
xtick={-20,0,20,40,60,80,100},
xticklabels={0.0,0.5,1.0,,,,},
y dir=reverse,
y grid style={darkgray176},
ymin=-0.5, ymax=99.5,
ytick pos=left,
ytick style={color=black},
ytick={-20,0,20,40,60,80,100},
yticklabels={0.0,0.5,1.0,,,,}
]
\addplot graphics [includegraphics cmd=\pgfimage,xmin=-0.5, xmax=99.5, ymin=99.5, ymax=-0.5] {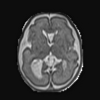};
\draw[red,very thin] (axis cs:33.08,25.63) rectangle (axis cs:57.48,50.62);

\nextgroupplot[
tick align=outside,
title={c) MIAL (3)},
x grid style={darkgray176},
xmin=-0.5, xmax=99.5,
xtick pos=right,
xtick style={color=black},
xtick={-20,0,20,40,60,80,100},
xticklabels={0.0,0.5,1.0,,,,},
y dir=reverse,
y grid style={darkgray176},
ymin=-0.5, ymax=99.5,
ytick pos=left,
ytick style={color=black},
ytick={-20,0,20,40,60,80,100},
yticklabels={0.0,0.5,1.0,,,,}
]
\addplot graphics [includegraphics cmd=\pgfimage,xmin=-0.5, xmax=99.5, ymin=99.5, ymax=-0.5] {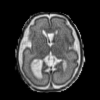};
\draw[red,very thin] (axis cs:33.08,25.63) rectangle (axis cs:57.48,50.62);

\nextgroupplot[
tick align=outside,
x grid style={darkgray176},
xmin=-0.5, xmax=99.5,
xtick pos=right,
xtick style={color=black},
xtick={-20,0,20,40,60,80,100},
xticklabels={0.0,0.5,1.0,,,,},
y dir=reverse,
ylabel={Sagittal},
ymin=-0.5, ymax=99.5,
ytick pos=left
]
\addplot graphics [includegraphics cmd=\pgfimage,xmin=-0.5, xmax=99.5, ymin=99.5, ymax=-0.5] {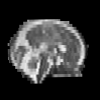};
\draw[red,very thin] (axis cs:42.026801223138,48.8892604304148) rectangle (axis cs:66.4287147351192,73.8721718831575);

\nextgroupplot[
tick align=outside,
x grid style={darkgray176},
xmin=-0.5, xmax=99.5,
xtick pos=right,
xtick style={color=black},
xtick={-20,0,20,40,60,80,100},
xticklabels={0.0,0.5,1.0,,,,},
y dir=reverse,
y grid style={darkgray176},
ymin=-0.5, ymax=99.5,
ytick pos=left,
ytick style={color=black},
ytick={-20,0,20,40,60,80,100},
yticklabels={0.0,0.5,1.0,,,,}
]
\addplot graphics [includegraphics cmd=\pgfimage,xmin=-0.5, xmax=99.5, ymin=99.5, ymax=-0.5] {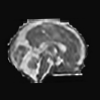};
\draw[red,very thin] (axis cs:42.026801223138,48.8892604304148) rectangle (axis cs:66.4287147351192,73.8721718831575);

\nextgroupplot[
tick align=outside,
x grid style={darkgray176},
xmin=-0.5, xmax=99.5,
xtick pos=right,
xtick style={color=black},
xtick={-20,0,20,40,60,80,100},
xticklabels={0.0,0.5,1.0,,,,},
y dir=reverse,
y grid style={darkgray176},
ymin=-0.5, ymax=99.5,
ytick pos=left,
ytick style={color=black},
ytick={-20,0,20,40,60,80,100},
yticklabels={0.0,0.5,1.0,,,,}
]
\addplot graphics [includegraphics cmd=\pgfimage,xmin=-0.5, xmax=99.5, ymin=99.5, ymax=-0.5] {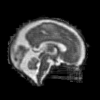};
\draw[red,very thin] (axis cs:42.026801223138,48.8892604304148) rectangle (axis cs:66.4287147351192,73.8721718831575);

\nextgroupplot[
tick align=outside,
x grid style={darkgray176},
xmin=-0.5, xmax=99.5,
xtick pos=right,
xtick style={color=black},
xtick={-20,0,20,40,60,80,100},
xticklabels={0.0,0.5,1.0,,,,},
y dir=reverse,
ylabel={Coronal},
ymin=-0.5, ymax=99.5,
ytick pos=left
]
\addplot graphics [includegraphics cmd=\pgfimage,xmin=-0.5, xmax=99.5, ymin=99.5, ymax=-0.5] {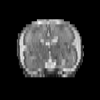};
\draw[red,very thin] (axis cs:42.026801223138,48.8892604304148) rectangle (axis cs:66.4287147351192,73.8721718831575);

\nextgroupplot[
tick align=outside,
x grid style={darkgray176},
xmin=-0.5, xmax=99.5,
xtick pos=right,
xtick style={color=black},
xtick={-20,0,20,40,60,80,100},
xticklabels={0.0,0.5,1.0,,,,},
y dir=reverse,
y grid style={darkgray176},
ymin=-0.5, ymax=99.5,
ytick pos=left,
ytick style={color=black},
ytick={-20,0,20,40,60,80,100},
yticklabels={0.0,0.5,1.0,,,,}
]
\addplot graphics [includegraphics cmd=\pgfimage,xmin=-0.5, xmax=99.5, ymin=99.5, ymax=-0.5] {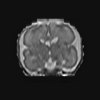};
\draw[red,very thin] (axis cs:42.026801223138,48.8892604304148) rectangle (axis cs:66.4287147351192,73.8721718831575);

\nextgroupplot[
tick align=outside,
x grid style={darkgray176},
xmin=-0.5, xmax=99.5,
xtick pos=right,
xtick style={color=black},
xtick={-20,0,20,40,60,80,100},
xticklabels={0.0,0.5,1.0,,,,},
y dir=reverse,
y grid style={darkgray176},
ymin=-0.5, ymax=99.5,
ytick pos=left,
ytick style={color=black},
ytick={-20,0,20,40,60,80,100},
yticklabels={0.0,0.5,1.0,,,,}
]
\addplot graphics [includegraphics cmd=\pgfimage,xmin=-0.5, xmax=99.5, ymin=99.5, ymax=-0.5] {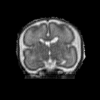};
\draw[red,very thin] (axis cs:42.026801223138,48.8892604304148) rectangle (axis cs:66.4287147351192,73.8721718831575);
\end{groupplot}

\end{tikzpicture}
\begin{tikzpicture}

\definecolor{darkgray176}{RGB}{176,176,176}

\begin{groupplot}[group style={group size=3 by 3,horizontal sep=\hs, vertical sep=\vs}, ticks=none, width=\imwidth,
title style={yshift=\titleshift},axis equal image]
\nextgroupplot[
tick align=outside,
title={a) LR},
x grid style={darkgray176},
xmin=33.08, xmax=57.48,
xtick pos=right,
xtick style={color=black},
xtick={40,45,50,55,60,65,70},
xticklabels={0.0,0.5,1.0,,,,},
y dir=reverse,
ymin=25.63, ymax=50.62,
ytick pos=left
]
\addplot graphics [includegraphics cmd=\pgfimage,xmin=-0.5, xmax=99.5, ymin=99.5, ymax=-0.5] {figs/anat_nomotion_1/anat_nomotion_1-000.png};

\nextgroupplot[
tick align=outside,
title={b) SAIR (1)},
x grid style={darkgray176},
xmin=33.08, xmax=57.48,
xtick pos=right,
xtick style={color=black},
xtick={40,45,50,55,60,65,70},
xticklabels={0.0,0.5,1.0,,,,},
y dir=reverse,
y grid style={darkgray176},
ymin=25.63, ymax=50.62,
ytick pos=left,
ytick style={color=black},
ytick={45,50,55,60,65,70,75},
yticklabels={0.0,0.5,1.0,,,,}
]
\addplot graphics [includegraphics cmd=\pgfimage,xmin=-0.5, xmax=99.5, ymin=99.5, ymax=-0.5] {figs/anat_nomotion_1/anat_nomotion_1-001.png};

\nextgroupplot[
tick align=outside,
title={c) MIAL (3)},
x grid style={darkgray176},
xmin=33.08, xmax=57.48,
xtick pos=right,
xtick style={color=black},
xtick={40,45,50,55,60,65,70},
xticklabels={0.0,0.5,1.0,,,,},
y dir=reverse,
y grid style={darkgray176},
ymin=25.63, ymax=50.62,
ytick pos=left,
ytick style={color=black},
ytick={45,50,55,60,65,70,75},
yticklabels={0.0,0.5,1.0,,,,}
]
\addplot graphics [includegraphics cmd=\pgfimage,xmin=-0.5, xmax=99.5, ymin=99.5, ymax=-0.5] {figs/anat_nomotion_1/anat_nomotion_1-002.png};

\nextgroupplot[
tick align=outside,
x grid style={darkgray176},
xmin=42.026801223138, xmax=66.4287147351192,
xtick pos=right,
xtick style={color=black},
xtick={40,45,50,55,60,65,70},
xticklabels={0.0,0.5,1.0,,,,},
y dir=reverse,
ymin=48.8892604304148, ymax=73.8721718831575,
ytick pos=left
]
\addplot graphics [includegraphics cmd=\pgfimage,xmin=-0.5, xmax=99.5, ymin=99.5, ymax=-0.5] {figs/anat_nomotion_1/anat_nomotion_1-003.png};

\nextgroupplot[
tick align=outside,
x grid style={darkgray176},
xmin=42.026801223138, xmax=66.4287147351192,
xtick pos=right,
xtick style={color=black},
xtick={40,45,50,55,60,65,70},
xticklabels={0.0,0.5,1.0,,,,},
y dir=reverse,
y grid style={darkgray176},
ymin=48.8892604304148, ymax=73.8721718831575,
ytick pos=left,
ytick style={color=black},
ytick={45,50,55,60,65,70,75},
yticklabels={0.0,0.5,1.0,,,,}
]
\addplot graphics [includegraphics cmd=\pgfimage,xmin=-0.5, xmax=99.5, ymin=99.5, ymax=-0.5] {figs/anat_nomotion_1/anat_nomotion_1-004.png};

\nextgroupplot[
tick align=outside,
x grid style={darkgray176},
xmin=42.026801223138, xmax=66.4287147351192,
xtick pos=right,
xtick style={color=black},
xtick={40,45,50,55,60,65,70},
xticklabels={0.0,0.5,1.0,,,,},
y dir=reverse,
y grid style={darkgray176},
ymin=48.8892604304148, ymax=73.8721718831575,
ytick pos=left,
ytick style={color=black},
ytick={45,50,55,60,65,70,75},
yticklabels={0.0,0.5,1.0,,,,}
]
\addplot graphics [includegraphics cmd=\pgfimage,xmin=-0.5, xmax=99.5, ymin=99.5, ymax=-0.5] {figs/anat_nomotion_1/anat_nomotion_1-005.png};

\nextgroupplot[
tick align=outside,
x grid style={darkgray176},
xmin=42.026801223138, xmax=66.4287147351192,
xtick pos=right,
xtick style={color=black},
xtick={40,45,50,55,60,65,70},
xticklabels={0.0,0.5,1.0,,,,},
y dir=reverse,
ymin=48.8892604304148, ymax=73.8721718831575,
ytick pos=left
]
\addplot graphics [includegraphics cmd=\pgfimage,xmin=-0.5, xmax=99.5, ymin=99.5, ymax=-0.5] {figs/anat_nomotion_1/anat_nomotion_1-006.png};

\nextgroupplot[
tick align=outside,
x grid style={darkgray176},
xmin=42.026801223138, xmax=66.4287147351192,
xtick pos=right,
xtick style={color=black},
xtick={40,45,50,55,60,65,70},
xticklabels={0.0,0.5,1.0,,,,},
y dir=reverse,
y grid style={darkgray176},
ymin=48.8892604304148, ymax=73.8721718831575,
ytick pos=left,
ytick style={color=black},
ytick={45,50,55,60,65,70,75},
yticklabels={0.0,0.5,1.0,,,,}
]
\addplot graphics [includegraphics cmd=\pgfimage,xmin=-0.5, xmax=99.5, ymin=99.5, ymax=-0.5] {figs/anat_nomotion_1/anat_nomotion_1-007.png};

\nextgroupplot[
tick align=outside,
x grid style={darkgray176},
xmin=42.026801223138, xmax=66.4287147351192,
xtick pos=right,
xtick style={color=black},
xtick={40,45,50,55,60,65,70},
xticklabels={0.0,0.5,1.0,,,,},
y dir=reverse,
y grid style={darkgray176},
ymin=48.8892604304148, ymax=73.8721718831575,
ytick pos=left,
ytick style={color=black},
ytick={45,50,55,60,65,70,75},
yticklabels={0.0,0.5,1.0,,,,}
]
\addplot graphics [includegraphics cmd=\pgfimage,xmin=-0.5, xmax=99.5, ymin=99.5, ymax=-0.5] {figs/anat_nomotion_1/anat_nomotion_1-008.png};
\end{groupplot}

\end{tikzpicture}

%% file: figs/anat_27_cor.tex
\def\imwidth{.185\textwidth}\def\hs{1pt}\def\vs{1pt}

\begin{tikzpicture}

\definecolor{darkgray176}{RGB}{176,176,176}

\begin{groupplot}[group style={group size=3 by 3,horizontal sep=\hs, vertical sep=\vs}, ticks=none, width=\imwidth,
title style={yshift=\titleshift}, axis equal image]
\nextgroupplot[
tick align=outside,
title={a) LR},
ylabel={Axial},
x grid style={darkgray176},
xmin=-0.5, xmax=99.5,
xtick pos=right,
xtick style={color=black},
xtick={-20,0,20,40,60,80,100},
xticklabels={0.0,0.5,1.0,,,,},
y dir=reverse,
ymin=-0.5, ymax=99.5,
ytick pos=left
]
\addplot graphics [includegraphics={keepaspectratio=true},xmin=-0.5, xmax=99.5, ymin=99.5, ymax=-0.5] {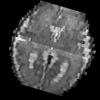};
\draw[red,very thin] (axis cs:42.088554539916,11.2789746765764) rectangle (axis cs:69.976455696466,39.7478737738878);

\nextgroupplot[
tick align=outside,
title={b) SAIR (1)},
x grid style={darkgray176},
xmin=-0.5, xmax=99.5,
xtick pos=right,
xtick style={color=black},
xtick={-20,0,20,40,60,80,100},
xticklabels={0.0,0.5,1.0,,,,},
y dir=reverse,
y grid style={darkgray176},
ymin=-0.5, ymax=99.5,
ytick pos=left,
ytick style={color=black},
ytick={-20,0,20,40,60,80,100},
yticklabels={0.0,0.5,1.0,,,,}
]
\addplot graphics [includegraphics cmd=\pgfimage,xmin=-0.5, xmax=99.5, ymin=99.5, ymax=-0.5] {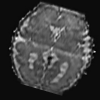};
\draw[red,very thin] (axis cs:42.088554539916,11.2789746765764) rectangle (axis cs:69.976455696466,39.7478737738878);

\nextgroupplot[
tick align=outside,
title={c) MIAL (8)},
x grid style={darkgray176},
xmin=-0.5, xmax=99.5,
xtick pos=right,
xtick style={color=black},
xtick={-20,0,20,40,60,80,100},
xticklabels={0.0,0.5,1.0,,,,},
y dir=reverse,
y grid style={darkgray176},
ymin=-0.5, ymax=99.5,
ytick pos=left,
ytick style={color=black},
ytick={-20,0,20,40,60,80,100},
yticklabels={0.0,0.5,1.0,,,,}
]
\addplot graphics [includegraphics cmd=\pgfimage,xmin=-0.5, xmax=99.5, ymin=99.5, ymax=-0.5] {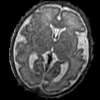};
\draw[red,very thin] (axis cs:42.088554539916,11.2789746765764) rectangle (axis cs:69.976455696466,39.7478737738878);

\nextgroupplot[
tick align=outside,
ylabel={Sagittal},
x grid style={darkgray176},
xmin=-0.5, xmax=99.5,
xtick pos=right,
xtick style={color=black},
xtick={-20,0,20,40,60,80,100},
xticklabels={0.0,0.5,1.0,,,,},
y dir=reverse,
ymin=-0.5, ymax=99.5,
ytick pos=left
]
\addplot graphics [includegraphics cmd=\pgfimage,xmin=-0.5, xmax=99.5, ymin=99.5, ymax=-0.5] {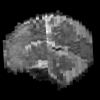};
\draw[red,very thin] (axis cs:42.088554539916,11.2789746765764) rectangle (axis cs:69.976455696466,39.7478737738878);

\nextgroupplot[
tick align=outside,
x grid style={darkgray176},
xmin=-0.5, xmax=99.5,
xtick pos=right,
xtick style={color=black},
xtick={-20,0,20,40,60,80,100},
xticklabels={0.0,0.5,1.0,,,,},
y dir=reverse,
y grid style={darkgray176},
ymin=-0.5, ymax=99.5,
ytick pos=left,
ytick style={color=black},
ytick={-20,0,20,40,60,80,100},
yticklabels={0.0,0.5,1.0,,,,}
]
\addplot graphics [includegraphics cmd=\pgfimage,xmin=-0.5, xmax=99.5, ymin=99.5, ymax=-0.5] {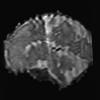};
\draw[red,very thin] (axis cs:42.088554539916,11.2789746765764) rectangle (axis cs:69.976455696466,39.7478737738878);

\nextgroupplot[
tick align=outside,
x grid style={darkgray176},
xmin=-0.5, xmax=99.5,
xtick pos=right,
xtick style={color=black},
xtick={-20,0,20,40,60,80,100},
xticklabels={0.0,0.5,1.0,,,,},
y dir=reverse,
y grid style={darkgray176},
ymin=-0.5, ymax=99.5,
ytick pos=left,
ytick style={color=black},
ytick={-20,0,20,40,60,80,100},
yticklabels={0.0,0.5,1.0,,,,}
]
\addplot graphics [includegraphics cmd=\pgfimage,xmin=-0.5, xmax=99.5, ymin=99.5, ymax=-0.5] {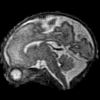};
\draw[red,very thin] (axis cs:42.088554539916,11.2789746765764) rectangle (axis cs:69.976455696466,39.7478737738878);

\nextgroupplot[
ylabel={Coronal},
tick align=outside,
x grid style={darkgray176},
xmin=-0.5, xmax=99.5,
xtick pos=right,
xtick style={color=black},
xtick={-20,0,20,40,60,80,100},
xticklabels={0.0,0.5,1.0,,,,},
y dir=reverse,
ymin=-0.5, ymax=99.5,
ytick pos=left
]
\addplot graphics [includegraphics cmd=\pgfimage,xmin=-0.5, xmax=99.5, ymin=99.5, ymax=-0.5] {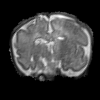};
\draw[red,very thin] (axis cs:42.088554539916,11.2789746765764) rectangle (axis cs:69.976455696466,39.7478737738878);

\nextgroupplot[
tick align=outside,
x grid style={darkgray176},
xmin=-0.5, xmax=99.5,
xtick pos=right,
xtick style={color=black},
xtick={-20,0,20,40,60,80,100},
xticklabels={0.0,0.5,1.0,,,,},
y dir=reverse,
y grid style={darkgray176},
ymin=-0.5, ymax=99.5,
ytick pos=left,
ytick style={color=black},
ytick={-20,0,20,40,60,80,100},
yticklabels={0.0,0.5,1.0,,,,}
]
\addplot graphics [includegraphics cmd=\pgfimage,xmin=-0.5, xmax=99.5, ymin=99.5, ymax=-0.5] {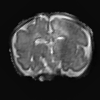};
\draw[red,very thin] (axis cs:42.088554539916,11.2789746765764) rectangle (axis cs:69.976455696466,39.7478737738878);

\nextgroupplot[
tick align=outside,
x grid style={darkgray176},
xmin=-0.5, xmax=99.5,
xtick pos=right,
xtick style={color=black},
xtick={-20,0,20,40,60,80,100},
xticklabels={0.0,0.5,1.0,,,,},
y dir=reverse,
y grid style={darkgray176},
ymin=-0.5, ymax=99.5,
ytick pos=left,
ytick style={color=black},
ytick={-20,0,20,40,60,80,100},
yticklabels={0.0,0.5,1.0,,,,}
]
\addplot graphics [includegraphics cmd=\pgfimage,xmin=-0.5, xmax=99.5, ymin=99.5, ymax=-0.5] {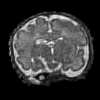};
\draw[red,very thin] (axis cs:42.088554539916,11.2789746765764) rectangle (axis cs:69.976455696466,39.7478737738878);
\end{groupplot}

\end{tikzpicture}
\begin{tikzpicture}

\definecolor{darkgray176}{RGB}{176,176,176}

\begin{groupplot}[group style={group size=3 by 3,horizontal sep=\hs, vertical sep=\vs}, ticks=none, width=\imwidth,
title style={yshift=\titleshift},axis equal image]
\nextgroupplot[
tick align=outside,
title={a) LR},
x grid style={darkgray176},
xmin=42.088554539916, xmax=69.976455696466,
xtick pos=right,
xtick style={color=black},
xtick={40,50,60,70},
xticklabels={0.0,0.5,1.0,},
y dir=reverse,
ymin=11.2789746765764, ymax=39.7478737738878,
ytick pos=left
]
\addplot graphics [includegraphics cmd=\pgfimage,xmin=-0.5, xmax=99.5, ymin=99.5, ymax=-0.5] {figs/anat_27_cor/anat_27_cor-000.png};

\nextgroupplot[
tick align=outside,
title={b) SAIR (1)},
x grid style={darkgray176},
xmin=42.088554539916, xmax=69.976455696466,
xtick pos=right,
xtick style={color=black},
xtick={40,50,60,70},
xticklabels={0.0,0.5,1.0,},
y dir=reverse,
y grid style={darkgray176},
ymin=11.2789746765764, ymax=39.7478737738878,
ytick pos=left,
ytick style={color=black},
ytick={10,15,20,25,30,35,40},
yticklabels={0.0,0.5,1.0,,,,}
]
\addplot graphics [includegraphics cmd=\pgfimage,xmin=-0.5, xmax=99.5, ymin=99.5, ymax=-0.5] {figs/anat_27_cor/anat_27_cor-001.png};

\nextgroupplot[
tick align=outside,
title={c) MIAL (8)},
x grid style={darkgray176},
xmin=42.088554539916, xmax=69.976455696466,
xtick pos=right,
xtick style={color=black},
xtick={40,50,60,70},
xticklabels={0.0,0.5,1.0,},
y dir=reverse,
y grid style={darkgray176},
ymin=11.2789746765764, ymax=39.7478737738878,
ytick pos=left,
ytick style={color=black},
ytick={10,15,20,25,30,35,40},
yticklabels={0.0,0.5,1.0,,,,}
]
\addplot graphics [includegraphics cmd=\pgfimage,xmin=-0.5, xmax=99.5, ymin=99.5, ymax=-0.5] {figs/anat_27_cor/anat_27_cor-002.png};

\nextgroupplot[
tick align=outside,
x grid style={darkgray176},
xmin=42.088554539916, xmax=69.976455696466,
xtick pos=right,
xtick style={color=black},
xtick={40,50,60,70},
xticklabels={0.0,0.5,1.0,},
y dir=reverse,
ymin=11.2789746765764, ymax=39.7478737738878,
ytick pos=left
]
\addplot graphics [includegraphics cmd=\pgfimage,xmin=-0.5, xmax=99.5, ymin=99.5, ymax=-0.5] {figs/anat_27_cor/anat_27_cor-003.png};

\nextgroupplot[
tick align=outside,
x grid style={darkgray176},
xmin=42.088554539916, xmax=69.976455696466,
xtick pos=right,
xtick style={color=black},
xtick={40,50,60,70},
xticklabels={0.0,0.5,1.0,},
y dir=reverse,
y grid style={darkgray176},
ymin=11.2789746765764, ymax=39.7478737738878,
ytick pos=left,
ytick style={color=black},
ytick={10,15,20,25,30,35,40},
yticklabels={0.0,0.5,1.0,,,,}
]
\addplot graphics [includegraphics cmd=\pgfimage,xmin=-0.5, xmax=99.5, ymin=99.5, ymax=-0.5] {figs/anat_27_cor/anat_27_cor-004.png};

\nextgroupplot[
tick align=outside,
x grid style={darkgray176},
xmin=42.088554539916, xmax=69.976455696466,
xtick pos=right,
xtick style={color=black},
xtick={40,50,60,70},
xticklabels={0.0,0.5,1.0,},
y dir=reverse,
y grid style={darkgray176},
ymin=11.2789746765764, ymax=39.7478737738878,
ytick pos=left,
ytick style={color=black},
ytick={10,15,20,25,30,35,40},
yticklabels={0.0,0.5,1.0,,,,}
]
\addplot graphics [includegraphics cmd=\pgfimage,xmin=-0.5, xmax=99.5, ymin=99.5, ymax=-0.5] {figs/anat_27_cor/anat_27_cor-005.png};

\nextgroupplot[
tick align=outside,
x grid style={darkgray176},
xmin=42.088554539916, xmax=69.976455696466,
xtick pos=right,
xtick style={color=black},
xtick={40,50,60,70},
xticklabels={0.0,0.5,1.0,},
y dir=reverse,
ymin=11.2789746765764, ymax=39.7478737738878,
ytick pos=left
]
\addplot graphics [includegraphics cmd=\pgfimage,xmin=-0.5, xmax=99.5, ymin=99.5, ymax=-0.5] {figs/anat_27_cor/anat_27_cor-006.png};

\nextgroupplot[
tick align=outside,
x grid style={darkgray176},
xmin=42.088554539916, xmax=69.976455696466,
xtick pos=right,
xtick style={color=black},
xtick={40,50,60,70},
xticklabels={0.0,0.5,1.0,},
y dir=reverse,
y grid style={darkgray176},
ymin=11.2789746765764, ymax=39.7478737738878,
ytick pos=left,
ytick style={color=black},
ytick={10,15,20,25,30,35,40},
yticklabels={0.0,0.5,1.0,,,,}
]
\addplot graphics [includegraphics cmd=\pgfimage,xmin=-0.5, xmax=99.5, ymin=99.5, ymax=-0.5] {figs/anat_27_cor/anat_27_cor-007.png};

\nextgroupplot[
tick align=outside,
x grid style={darkgray176},
xmin=42.088554539916, xmax=69.976455696466,
xtick pos=right,
xtick style={color=black},
xtick={40,50,60,70},
xticklabels={0.0,0.5,1.0,},
y dir=reverse,
y grid style={darkgray176},
ymin=11.2789746765764, ymax=39.7478737738878,
ytick pos=left,
ytick style={color=black},
ytick={10,15,20,25,30,35,40},
yticklabels={0.0,0.5,1.0,,,,}
]
\addplot graphics [includegraphics cmd=\pgfimage,xmin=-0.5, xmax=99.5, ymin=99.5, ymax=-0.5] {figs/anat_27_cor/anat_27_cor-008.png};
\end{groupplot}

\end{tikzpicture}

%% file: secs/discussion.tex
In this work, we present the first study of a single-volume self-supervised superresolution method for the reconstruction of T2-weighted magnetic-resonance images (MRI) of the fetal brain compared to more conventional reconstruction techniques such as MIALSRTK that combine multiple orthogonal low-resolution volumes. We exemplify the accuracy of our proposed single-acquisition isotropic resolution method on synthetic data without motion. Such simulations rely on a numerical phantom and are highly valuable to evaluate the proposed method. They provide not only more realistic LR images than the simplified MRI model commonly used, but also high-resolution isotropic 
ground-truth images. As proof of concept, we also show the applicability of SAIR on real clinical acquisitions with little amplitude of fetal movements.
The resulting single-volume superresolution method will be very useful to replace the interpolation methods currently used to upsample the reference anatomy needed in the motion-estimation step.
In addition, the reconstruction of a high-resolution volume of the fetal brain from one single series would be of paramount interest in the perspective of a clinical translation to minimize the acquisition time in this cohort of sensitive subjects. Further development will focus on integrating SAIR with motion-estimation methods, for instance, based on an age-matched HR simulated volume.